\documentclass[floatfix,twocolumn,showpacs,preprintnumbers,amsmath,amssymb,pra,superscriptaddress,longbibliography]{revtex4-1}
\usepackage{color}
\usepackage[colorlinks=true,linkcolor=blue,urlcolor=blue,citecolor=blue]{hyperref}
\usepackage{mathtools}
\usepackage{graphicx}
\usepackage{dcolumn}
\usepackage{array}
\usepackage{lipsum}
\usepackage{bm}
\usepackage{subfigure}
\usepackage{amssymb}
\usepackage{multirow}
\usepackage{tabularx}
\usepackage{amsmath}
\usepackage{braket}
\graphicspath{{plots/}}
 \usepackage{lipsum}
\usepackage{mathrsfs}
	
\newcommand{\beq}{\begin{equation}}
\newcommand{\eeq}{\end{equation}}
\newcommand{\bea}{\begin{eqnarray}}
\newcommand{\eea}{\end{eqnarray}}

\begin{document}
\title{Energy response and spatial alignment of the perturbed electron gas}
\author{Tobias Dornheim}
\email{t.dornheim@hzdr.de}
\affiliation{Center for Advanced Systems Understanding (CASUS), D-02826 G\"orlitz, Germany}
\affiliation{Helmholtz-Zentrum Dresden-Rossendorf (HZDR), D-01328 Dresden, Germany}
\author{Panagiotis Tolias}
\affiliation{Space and Plasma Physics, Royal Institute of Technology (KTH), Stockholm, SE-100 44, Sweden}
\author{Zhandos A.~Moldabekov}
\affiliation{Center for Advanced Systems Understanding (CASUS), D-02826 G\"orlitz, Germany}
\affiliation{Helmholtz-Zentrum Dresden-Rossendorf (HZDR), D-01328 Dresden, Germany}
\author{Jan Vorberger}
\affiliation{Helmholtz-Zentrum Dresden-Rossendorf (HZDR), D-01328 Dresden, Germany}
\begin{abstract}
We present extensive new \emph{ab initio} path integral Monte Carlo (PIMC) simulations of the harmonically perturbed uniform electron gas (UEG) for different densities and temperatures. This allows us to study the linear response of the UEG with respect to different contributions to the total energy for different wave numbers. We find that the induced change in the interaction energy exhibits a non-monotonic behaviour, and becomes negative for intermediate wave numbers. This effect is strongly dependent on the coupling strength and can be traced back to the spatial alignment of electrons introduced in earlier works [T.~Dornheim \emph{et al.}, Communications Physics \textbf{5}, 304 (2022)]. The observed quadratic dependence on the perturbation amplitude in the limit of weak perturbations and the quartic dependence of the perturbation amplitude corrections are consistent with linear and non-linear versions of the density stifness theorem. All PIMC simulation results are freely available online and can be used to benchmark new methods, or as input for other calculations.
\end{abstract}
\maketitle

\section{Introduction\label{sec:introduction}}

The uniform electron gas (UEG)~\cite{loos,quantum_theory,review} constitutes one of the most actively investigated model systems in the literature. Despite its apparent simplicity, accurate knowledge of its properties requires state-of-the-art numerical simulations. In this regard, the most successful approach is offered by various quantum Monte Carlo (QMC) techniques~\cite{anderson2007quantum,Foulkes_RMP_2001}. In particular, the availability of highly accurate QMC results in the electronic ground state~\cite{Ceperley_Alder_PRL_1980,Spink_PRB_2013,moroni,moroni2,Ortiz_PRB_1994,Ortiz_PRL_1999} (at temperature $T=0$) and their subsequent use as input for the analytical parametrization of various UEG properties~\cite{vwn,Perdew_Zunger_PRB_1981,cdop,Gori-Giorgi_PRB_2000,farid} has been of paramount importance serving as the basis for the arguably unrivaled success of density functional theory (DFT)~\cite{Jones_RMP_2015}. In fact, the UEG ground state remains an active topic, and new QMC results have been presented over the last years both in the limits of high density~\cite{Shepherd_JCP_2012,https://doi.org/10.48550/arxiv.2209.10227} and strong coupling~\cite{Holzmann_PRL_2020,Azadi_PRB_2022}.

In addition, the recent surge of interest in matter at extreme temperatures ($T=10^3-10^8\,$K) and pressures ($P=1-10^4\,$MBar) as they occur in astrophysical objects (e.g.~giant planet interiors~\cite{Benuzzi_Mounaix_2014} or brown dwarfs~\cite{becker}) and cutting-edge technological applications (e.g.~inertial confinement fusion experiments~\cite{hu_ICF,Betti2016} and the discovery of novel materials~\cite{Kraus2016,Kraus2017,Lazicki2021}) has boosted efforts on the QMC simulation of the UEG at finite temperatures~\cite{Brown_PRL_2013,Brown_PRB_2013,Schoof_PRL_2015,dornheim_prl,groth_prl,Dornheim_JCP_2015,Malone_JCP_2015,Malone_PRL_2016,Joonho_JCP_2021,Yilmaz_JCP_2020,dornheim_pre,groth_jcp,Dornheim_PRL_2020,dornheim_ML,Dornheim_permutation_cycles,dornheim_HEDP,Dornheim_HEDP_2022,dornheim_electron_liquid,Dornheim_PRB_nk_2021,Dornheim_PRE_2021,Dornheim_JCP_2021,Filinov_PRE_2015,Filinov_CPP_2021}, see Refs.~\cite{review},\cite{Dornheim_POP_2017} for overviews of some developments. These \emph{warm dense matter} (WDM) conditions are typically characterised in terms of 1) the density parameter $r_s=\overline{r}/a_\textnormal{B}$, with $\overline{r}$ the mean interparticle distance and $a_\textnormal{B}$ the first Bohr radius, 2) the degeneracy temperature~\cite{Ott2018} $\Theta=k_\textnormal{B}T/E_\textnormal{F}$, with $E_\textnormal{F}$ the usual Fermi energy~\cite{quantum_theory}. From a physical perspective, the condition $r_s\sim\Theta\sim1$ in the WDM regime implies a nontrivial interplay of Coulomb coupling, quantum degeneracy and thermal excitation effects, which renders the accurate theoretical description of WDM a most formidable challenge~\cite{wdm_book,new_POP}.
The most pressing obstacle is given by the notorious fermion sign problem~\cite{Loh_PRB_1990,troyer,dornheim_sign_problem,Dornheim_JPA_2021}, which leads to an exponential increase in the required compute time upon increasing the system size $N$, or upon decreasing the temperature $T$. Nevertheless, considerable efforts by different groups have led to the first accurate parametrizations of the exchange--correlation free energy of the UEG~\cite{ksdt,groth_prl,review} that cover the entire relevant WDM parameter space.

A particularly important property of the UEG concerns its response to external perturbations~\cite{Dornheim_review}. Such information constitutes valuable input for various applications, including the construction of advanced exchange--correlation functionals for DFT~\cite{Patrick_JCP_2015,pribram} and the computation of electronically screened effective ion--ion potentials~\cite{ceperley_potential,zhandos_cpp17,zhandos1}.\,At the ground state, the first highly accurate QMC results for the static linear density response function have been presented in pioneering works by Moroni \emph{et al.}~\cite{moroni,moroni2} and subsequently parametrized by Corradini \emph{et al.}~\cite{cdop}. These data have been further substantiated and complemented by recent diagrammatic QMC results~\cite{Chen2019}. The original idea from Ref.~\cite{moroni} has also been adapted to the finite temperature UEG via different path integral Monte Carlo (PIMC) methods in Refs.~\cite{dornheim_pre,groth_jcp}. This has been followed by extensive standard PIMC studies of the linear density response of the warm dense UEG based on imaginary-time density--density correlation functions by Dornheim and co-workers~\cite{dornheim_ML,dornheim_HEDP,dornheim_electron_liquid}, which have subsequently been confirmed in the recent Ref.~\cite{PhysRevB.106.L081126}. In particular, the PIMC results have been used as the basis for a neural-net representation of the static electronic local field correction, and have become available as an analytical representation~\cite{Dornheim_PRB_ESA_2021} that incorporates different limits~\cite{Dornheim_PRL_2020_ESA}.
In the mean time, these investigations have been extended to the nonlinear regime~\cite{Dornheim_PRL_2020,Dornheim_PRR_2021,Dornheim_CPP_2022,Dornheim_JPSJ_2021,Moldabekov_SciRep_2022,Moldabekov_JCTC_2022}, which, interestingly, is connected to the higher-order imaginary-time correlation functions~\cite{Dornheim_JCP_ITCF_2021}.

A further interesting dimension for the study of the linear density response of the UEG is given by its dependence on the perturbation frequency $\omega$. In this regard, a key property is given by the dynamic structure factor $S(\mathbf{q},\omega)$, where $\mathbf{q}$ denotes the wave vector. In particular, $S(\mathbf{q},\omega)$ plays a central role in the interpretation of X-ray Thomson scattering (XRTS)~\cite{siegfried_review,kraus_xrts,Dornheim_T_2022,Dornheim_T2_2022}, which constitutes an important method of diagnostics for WDM experiments~\cite{falk_wdm}. Here a central problem is given by the analytic continuation of PIMC results, which are usually limited to the imaginary time, to the real-frequency domain---a well-known, ill-posed problem~\cite{JARRELL1996133,PhysRevB.94.245140,Goulko_PRB_2017}. Consequently, the first highly accurate results for the dynamic structure factor of the UEG have been presented only recently based on the constrained stochastic sampling of the dynamic local field correction~\cite{dornheim_dynamic,dynamic_folgepaper,Dornheim_PRE_2020}.
Interestingly, it has been found that the exact incorporation of dynamic exchange--correlation effects leads to a red-shift compared to the random phase approximation (RPA) at metallic density, and a \emph{roton-type} minimum in the position of the maximum of $S(\mathbf{q},\omega)$, $\omega(q)$, around twice the Fermi wave number $q_\textnormal{F}$ at strong coupling. For completeness, 
we note that these findings are in qualitative agreement to earlier predictions by Takada \emph{et al.}~\cite{Takada_PRB_2016,Takada_PRL_2002} in the ground-state limit based on an approximate many-body treatment of the UEG,
as well as the very recent investigation based on the Bethe-Salpeter equation using vertex corrections presented in Ref.~\cite{Koskelo}.
Both the roton feature and the exchange--correlation induced red-shift have subsequently been explained by the alignment of pairs of electrons~\cite{Dornheim_Nature_2022,Dornheim_Force_2022}, which leads to an effective reduction in the interaction energy when the wave length $\lambda=2\pi/q$ of such a density fluctuation is comparable to the average interparticle distance $d$.

The present work is devoted to a further exploration of the linear density response of the warm dense UEG, in general, and of the physical validity of the pair alignment idea presented in Ref.~\cite{Dornheim_Nature_2022}, in particular. To this end, we present new, extensive \emph{ab initio} PIMC results for the static density response properties of the UEG, with a particular emphasis on the respective change of different energies. Indeed, we find that the interaction energy $W$ is reduced when $\lambda\sim d$, as it has been predicted~\cite{Dornheim_Nature_2022}. This effect becomes more pronounced in the low-density regime, when the system approaches a strongly coupled electron liquid, and is reduced upon increasing the temperature. For a more qualitative insight into these observations, we make use of the stiffness theorem~\cite{quantum_theory} (see Eq.~(\ref{eq:stiffness}) below), which links the change in the free energy $F$ due to an external perturbation to the aforementioned static density response function $\chi(\mathbf{q})$. A generalized nonlinear formulation is also discussed.

The paper is organized as follows. In Sec.~\ref{sec:theory}, the theoretical background is introduced, namely a brief discussion of the UEG~(\ref{sec:UEG}), the PIMC method~(\ref{sec:PIMC}), linear response theory (\ref{sec:LRT}) and the density stiffness theorem (\ref{sec:DST}). Sec.~\ref{sec:results} is devoted to the presentation and analysis of our new simulation results, starting with the strongly coupled electron regime in Sec.~\ref{sec:electron_liquid}. In addition, we consider the effects of temperature and density in Secs.~\ref{sec:temperature} and \ref{sec:coupling}, respectively. The manuscript is concluded with a summary and outlook in Sec.~\ref{sec:summary}. Nonlinear generalizations are explored in the appendices \ref{sec:appendixA} and \ref{sec:appendixB}.

\section{Theory\label{sec:theory}}

Note that we assume Hartree atomic units throughout this work.

\subsection{Uniform electron gas\label{sec:UEG}}

The UEG, sometimes referred to as \emph{jellium} in the literature, constitutes the quantum version of the classical one-component plasma (OCP) model~\cite{Baus_Hansen_OCP,plasma2,OCP_bridge_2022}. In particular, we assume $N$ electrons of (negative) unit charge that are immersed in a cubic volume $\Omega=L^3$, where $L$ is the length of the periodic simulation cell. The corresponding Hamiltonian can be expressed as
\begin{eqnarray}\label{eq:Hamiltonian}
\hat{H}_\textnormal{UEG} = -\frac{1}{2}\sum_{l=1}^N \nabla_l^2 + \sum_{l=1}^N \sum_{k>l}^N W_\textnormal{E}(\hat{\mathbf{r}}_l,\hat{\mathbf{r}}_k)\ ,
\end{eqnarray}
where the first RHS term corresponds to the usual kinetic energy $\hat{K}$, and where the Ewald potential $W_\textnormal{E}(\mathbf{r}_a,\mathbf{r}_b)$ contains the sum over all periodic images, the interaction between the electrons and the uniform positive background, as well as the Madelung constant; see the excellent discussion by Fraser \emph{et al.}~\cite{Fraser_PRB_1996} for more details.

The thermodynamic states of the UEG are fully determined by three dimensionless parameters. The Wigner-Seitz radius $r_s$ serves as the quantum coupling parameter. Thus, the UEG attains the limit of an ideal (i.e., noninteracting) Fermi gas for $r_s\to0$ and forms a Wigner crystal~\cite{Wigner_PhysRev_1934,Azadi_Wigner_2022,Drummond_PRB_Wigner_2004} for $r_s\sim100$. The temperature parameter $\Theta$ indicates the degree of quantum degeneracy, with $\Theta=0$ the fully degenerate ground state limit, and $\Theta\gg1$ corresponding to the classical limit~\cite{Dornheim_HEDP_2022}. The WDM regime is characterised by $\Theta\sim r_s\sim1$, which indicates partial degeneracy and moderate coupling. As a consequence, no reliable expansions around the classical limit, the ground-state limit, the non-interacting limit or the perfect crystal limit are possible, and an accurate theoretical description requires state-of-the-art numerical methods~\cite{wdm_book,new_POP}. The third parameter is given by the spin-polarization $\xi=(N^\uparrow-N^\downarrow)/N$, where $N^\uparrow$ and $N^\downarrow$ denote the majority and minority spin-orientations, with $N=N^\uparrow+N^\downarrow$. We restrict ourselves to the fully unpolarized (i.e., paramagnetic) limit of $\xi=0$ throughout this work.

\subsection{Path integral Monte Carlo\label{sec:PIMC}}

In order to acquire highly accurate results for the density response of the UEG, we employ the \emph{ab initio} PIMC method~\cite{cep,Berne_JCP_1982,Takahashi_Imada_PIMC_1984} without any restrictions on the nodal structure of the thermal density matrix~\cite{Ceperley1991}. Therefore, our simulations are computationally expensive due to the fermion sign problem~\cite{dornheim_sign_problem}, but exact within the given Monte Carlo error bars. We estimate a total simulation time of $\mathcal{O}\left(10^6\right)$ CPUh for the present study. 

The starting point of the PIMC method is to express the canonical (i.e., volume $\Omega$, number density $n=N/\Omega$ and inverse temperature $\beta=1/T$ are constant) partition function in coordinate space as  
\begin{widetext} 
\begin{eqnarray}\label{eq:Z}
Z_{\beta,N,\Omega} &=& \frac{1}{N^\uparrow! N^\downarrow!} \sum_{\sigma^\uparrow\in S_N^\uparrow} \sum_{\sigma^\downarrow\in S_N^\downarrow} \textnormal{sgn}(\sigma^\uparrow,\sigma^\downarrow) \int_\Omega d\mathbf{R} \bra{\mathbf{R}} e^{-\beta\hat H} \ket{\hat{\pi}_{\sigma^\uparrow}\hat{\pi}_{\sigma^\downarrow}\mathbf{R}}\ ,
\end{eqnarray}\end{widetext}
with the meta variable $\mathbf{R}=(\mathbf{r}_1,\dots,\mathbf{r}_N)^T$ containing the coordinates of all $N$ electrons.
The anti-symmetry of $Z$ with respect to the exchange of two electrons of the same spin-orientation is taken into account by the sums over all possible permutation elements $\sigma^i$ ($i\in\{\uparrow,\downarrow\}$) of the respective permutation group $S_N^i$, and the corresponding permutation operators $\hat{\pi}_{\sigma^i}$.
Unfortunately, Eq.~(\ref{eq:Z}) is of no direct practical value, since the matrix elements of the density operator $\hat\rho=e^{-\beta\hat H}$ cannot be readily evaluated; this is an immediate consequence of the fact that the kinetic and potential contributions to the full Hamiltonian operator $\hat H=\hat K + \hat V$ do not commute~\cite{kleinert2009path}.
To work around this limitation, we employ the exact semi-group property of the density operator
\begin{eqnarray}\label{eq:group}
e^{-\beta\hat{H}} = \left(e^{-\epsilon\hat{H}} \right)^P \ ,
\end{eqnarray}
where we have defined the reduced inverse temperature $\epsilon=\beta/P$. Inserting Eq.~(\ref{eq:group}) into (\ref{eq:Z}) leads to
\begin{widetext}
\begin{eqnarray}\label{eq:Z_modified}
Z_{\beta,N,\Omega} &=& \frac{1}{N^\uparrow! N^\downarrow!} \sum_{\sigma^\uparrow\in S_N} \sum_{\sigma^\downarrow\in S_N} \textnormal{sgn}(\sigma^\uparrow,\sigma^\downarrow)\int_\Omega d\mathbf{R}_0\dots d\mathbf{R}_{P-1}
\bra{\mathbf{R}_0}e^{-\epsilon\hat H}\ket{\mathbf{R}_1}  \dots 
\bra{\mathbf{R}_{P-1}} e^{-\epsilon\hat H} \ket{\hat{\pi}_{\sigma^\uparrow}\hat{\pi}_{\sigma^\downarrow}\mathbf{R}_0}\ , \quad
\end{eqnarray}
\end{widetext}
where the single matrix element of the density operator $e^{-\beta\hat H}$ has been replaced by a product of $P$ density operators at $P$-times the original temperature. In the limit of large $P$, we can introduce a suitable high-temperature approximation to the density matrix, and the corresponding factorization error can be made arbitrarily small by increasing $P$. In the present work, we use the \emph{primitive factorization}
\begin{eqnarray}
e^{-\epsilon \hat H} \approx e^{-\epsilon \hat K} e^{-\epsilon \hat V}\ ,
\end{eqnarray}
and the corresponding factorization error vanishes as $\mathcal{O}\left(P^{-2}\right)$. We note that higher-order factorizations have been considered in the literature~\cite{sakkos_JCP_2009,brualla_JCP_2004,Zillich_JCP_2010,Dornheim_NJP_2015,Chin_PRE_2015,Dornheim_CPP_2019}, but they are not required for the temperatures that are relevant for the present investigation. In practice, we find that $P=200$ is fully sufficient to ensure convergence; see the Supplemental Material of Ref.~\cite{Dornheim_PRL_2020} for a corresponding convergence study.

Finally, we note that all the PIMC results that are presented in this work have been obtained based on the extended ensemble scheme discussed in Ref.~\cite{Dornheim_PRB_nk_2021}, which constitutes a canonical version of the worm algorithm by Boninsegni \emph{et al.}~\cite{boninsegni1,boninsegni2}.

\subsection{Linear density response theory\label{sec:LRT}}

For the study of the linear UEG density response, we follow the original idea by Moroni \emph{et al.}~\cite{moroni,moroni2} and Bowen \emph{et al.}~\cite{bowen2}, and consider the modified Hamiltonian
\begin{eqnarray}\label{eq:Hamiltonian_modified}
\hat H = \hat{H}_\textnormal{UEG} + 2A\sum_{l=1}^N\textnormal{cos}\left(\mathbf{q}\cdot\hat{\mathbf{r}}_l\right)\ ,
\end{eqnarray}
with the last term representing a single external harmonic static perturbation; the application of multiple perturbations is in principle possible and it has been investigated in the linear~\cite{dornheim_pre} and the non-linear~\cite{Dornheim_CPP_2022} response regimes. In addition, we note that the basic idea of Eq.~(\ref{eq:Hamiltonian_modified}) is in no way unique to the UEG, and has successfully been used to study the static density response of a gamut of systems including charged bosons~\cite{Sugiyama_PRB_1994}, neutron matter~\cite{PhysRevLett.116.152501}, and warm dense hydrogen~\cite{Bohme_PRL_2022,Bohme_PRE_2023}. Furthermore, it can be adapted to simulation methods other than QMC, such as the approximate, though much less demanding computationally, DFT approach~\cite{Moldabekov_JCTC_2022,doi:10.1063/5.0062325}.

To measure the density response of the thus perturbed UEG, we compute the thermal expectation value of the density in reciprocal space
\begin{eqnarray}\label{eq:rho}
\braket{\hat\rho_\mathbf{k}}_{q,A} = \frac{1}{\Omega} \left< \sum_{l=1}^N e^{-i\mathbf{k}\cdot\hat{\mathbf{r}}_l} \right>_{q,A} \ , 
\end{eqnarray}
where $\mathbf{q}$ and $\mathbf{k}$ denote the wave vectors of the external perturbation and of the UEG systemic response, respectively. Throughout this work, we restrict ourselves to the linear response regime, which is valid when the perturbation amplitude $A$ is sufficiently small. In this case, it simply holds
\begin{eqnarray}\label{eq:LRT}
\braket{\hat\rho_\mathbf{k}}_{q,A} = \delta_{\mathbf{q},\mathbf{k}} \chi(\mathbf{q}) A\ ,
\end{eqnarray}
where the static linear response function is typically expressed as~\cite{kugler1}
\begin{eqnarray}\label{eq:LFC}
\chi(\mathbf{q}) = \frac{\chi_0(\mathbf{q})}{1-\frac{4\pi}{q^2}\left[ 1-G(\mathbf{q})\right]\chi_0(\mathbf{q})}\ ,
\end{eqnarray}
with $\chi_0(\mathbf{q})$ the linear density response of the ideal Fermi gas (also known as finite-$T$ Lindhard function in the literature~\cite{quantum_theory}), which can be readily computed. In addition, the static local field correction $G(\mathbf{q})$ contains the full information about electronic exchange--correlation effects and constitutes the static limit of the dynamic local field correction $G(\mathbf{q},\omega)$, which is the fundamental quantity in all dielectric theories, see e.g., Refs.~\cite{stls_original,vs_original,stls,stls2,stolzmann,tanaka_hnc,Tanaka_CPP_2017,arora,Tolias_JCP_2021,castello2021classical,Tolias_arxiv_2023}. Therefore, it constitutes key input also for a variety of applications including the construction of effective potentials~\cite{ceperley_potential,zhandos1}, quantum hydrodynamics~\cite{doi:10.1063/1.5003910,Diaw2017,Moldabekov_SciPost_2022} and the interpretation of XRTS experiments~\cite{siegfried_review,Fortmann_PRE_2010,kraus_xrts,Preston_APL_2019}. 

For completeness, we note that the static linear density response function $\chi(\mathbf{q})$ can also be estimated from PIMC simulations of the unperturbed system via the imaginary-time version of the fluctuation--dissipation theorem,
\begin{eqnarray}\label{eq:static_chi}
\chi(\mathbf{q}) = -n\int_0^\beta \textnormal{d}\tau\ F(\mathbf{q},\tau) \quad ;
\end{eqnarray}
see the appendix of Ref.~\cite{Dornheim_insight_2022} for an accessible derivation. Here $F(\mathbf{q},\tau)$ denotes the imaginary-time density--density correlation function that can be viewed as the imaginary-time version of the usual intermediate scattering function~\cite{siegfried_review}. Recent, extensive, discussions of the physical properties of $F(\mathbf{q},\tau)$ are available in Refs.~\cite{Dornheim_insight_2022,Dornheim_review,Dornheim_PTR_2022}.

\subsection{Density stiffness theorems\label{sec:DST}}

The general form of the ground state stiffness theorem connects the energy change associated with an external field that is applied to the UEG and alters the expectation value of an observable $\hat{\mathcal{O}}$ (originally zero) with the static response function that is associated with the same observable $\chi_{\mathcal{O}\mathcal{O}}(0)$~\cite{quantum_theory}. The proof involves a constrained minimization problem and it is based on linear response theory. In absence of spatial variations, one has
\begin{equation}\label{eq:generalgroundstatestiffness}
E_{\mathrm{UEG}}^{\mathcal{O}}-E_{\mathrm{UEG}}=\frac{1}{2}\frac{\mathcal{O}^2}{\chi_{\mathcal{O}\mathcal{O}}(0)}\,.
\end{equation}
The most standard form is the density stiffness theorem, see for instance the derivation of the compressibility sum rule~\cite{quantum_theory}. Consideration of the perturbed Hamiltonian of Eq.~(\ref{eq:Hamiltonian_modified}) and the inclusion of spatial variations, ultimately lead to the \emph{linear ground state density stiffness theorem}
\begin{equation}\label{eq:densitygroundstatestiffness}
E_{\mathrm{UEG}}^{\boldsymbol{q},A}-E_{\mathrm{UEG}}=-nA^2\chi(\boldsymbol{q})\,.
\end{equation}

The finite temperature density stiffness theorem relates the free energy difference between the harmonically perturbed electron gas and the UEG to the static density response. To our knowledge, the finite-T version has been sparsely discussed in the literature. It has been derived from a density functional perspective by minimizing the intrinsic free energy of the perturbed system with respect to the linear density perturbation, see the appendix C of Ref.~\cite{doi:10.1063/1.5003910}. An alternative derivation is also viable that utilizes the thermal density operator in the Hamiltonian eigenstate basis, employs the FT Hellmann-Feynman theorem~\cite{pons_2020} and, in all its other aspects, mirrors the well-known zero-temperature limit derivation~\cite{quantum_theory}. The \emph{linear finite temperature density stiffness theorem} reads as~\cite{doi:10.1063/1.5003910}
\begin{equation}\label{eq:stiffness}
F_{\mathrm{UEG}}^{\boldsymbol{q},A}-F_{\mathrm{UEG}}=-nA^2\chi(\mathbf{q})\,.
\end{equation}
In what follows, for a thermodynamic quantity of interest $Q$, we introduce the notation $\Delta{Q}=Q_{\mathrm{UEG}}^{\boldsymbol{q},A}-Q_{\mathrm{UEG}}$, with $Q_{\mathrm{UEG}}$ the UEG value and $Q_{\mathrm{UEG}}^{\boldsymbol{q},A}$ the Eq.~(\ref{eq:Hamiltonian_modified})-perturbed electron gas value. Thus, the above can be conveniently rewritten as
\begin{equation}\label{eq:stiffnessfree}
\Delta{F}=-\frac{3A^2}{4\pi{r}_{\mathrm{s}}^3}\chi(\boldsymbol{q};r_{\mathrm{s}},\Theta)\,.
\end{equation}
From Eq.~(\ref{eq:stiffnessfree}), it directly follows that it holds $\Delta Q\sim A^2$ for any thermodynamic observable.
For the internal energy, $F=E-TS$ combined with a first order thermodynamic entropy $S$ expression~\cite{review,groth_prl,ksdt,status} leads to\begin{equation}\label{eq:stiffnessinternal}
\Delta{E}=\frac{3A^2}{4\pi{r}_{\mathrm{s}}^3}\left(\Theta\frac{\partial}{\partial{\Theta}}-1\right)\chi(\boldsymbol{q};r_{\mathrm{s}},\Theta)\,.
\end{equation}
Moreover, we consider the external potential energy,  naturally zero for the UEG. In the limit of a vanishing perturbation strength, it is given by $V_{\mathrm{UEG}}^{\boldsymbol{q},A}=2nA^2\chi(\mathbf{q})$ (see the Appendix~\ref{sec:appendixA}) or equivalently by
\begin{eqnarray}\label{eq:stiffnessexternal}
V_{\mathrm{UEG}}^{\boldsymbol{q},A}=\frac{3A^2}{2\pi{r}_{\mathrm{s}}^3}\chi(\boldsymbol{q};r_{\mathrm{s}},\Theta)\,,
\end{eqnarray}
Finally, we mention the full nonlinear expression for the external potential energy (see the Appendix~\ref{sec:appendixA})
\begin{equation}\label{eq:stiffnessexternalnonlinear}
V_{\mathrm{UEG}}^{\boldsymbol{q},A}=2n\sum_{m=1}^{\infty}A^{2m}\chi^{(1,2m-1)}(\boldsymbol{q})\,,
\end{equation}
and the nonlinear finite temperature density stiffness theorem (see the Appendix~\ref{sec:appendixB})
\begin{equation}\label{eq:stiffnessfreenonlinear}
\Delta{F}=-n\sum_{m=1}^{\infty}\frac{2m-1}{m}A^{2m}\chi^{(1,2m-1)}(\boldsymbol{q})\,.
\end{equation}

Based on the non-linear generalizations of Eqs.~(\ref{eq:stiffnessexternalnonlinear},\ref{eq:stiffnessfreenonlinear}), the non-linear density stiffness theorem can now be combined with the standard thermodynamic relations to lead to more general expressions for the change in the different energy components as functionals of the non-linear density response functions. In this manner, the non-linear generalization of Eq.~(\ref{eq:stiffnessinternal}) can be straightforwardly obtained. Their general form reads as
\begin{eqnarray}\label{eq:energy_fit_general}
\Delta{Q}=\sum_{m=1}^{\infty}\hat{\gamma}_{\mathrm{Q},m}(r_{\mathrm{s}},\Theta)A^{2m}\chi^{(1,2m-1)}(\boldsymbol{q};r_{\mathrm{s}},\Theta),
\end{eqnarray}
where $Q=F,E,W,K,V_{\mathrm{ext}}$ with $W$ the interaction energy and $K$ the kinetic energy. We note that the form of the zero-order or first-order thermodynamic differential operator $\hat{\gamma}_{\mathrm{Q},m}(r_{\mathrm{s}},\Theta)$ depends on the energy component $Q$ but not on the non-linear order $m$ with the exception of the proportionality constant. For the values of $A$ considered herein, the limit of small but non-negligible perturbation strengths applies. Thus, the energy component changes can be accurately described by the first two terms of the series expansion. Therefore, we have
\begin{align}\label{eq:energy_fit_cubic}
\Delta{Q}&\simeq\hat{\gamma}_{\mathrm{Q},1}(r_{\mathrm{s}},\Theta)\chi^{(1,1)}(\boldsymbol{q};r_{\mathrm{s}},\Theta)A^{2}+\nonumber\\&\quad\,\,\,\hat{\gamma}_{\mathrm{Q},2}(r_{\mathrm{s}},\Theta)\chi^{(1,3)}(\boldsymbol{q};r_{\mathrm{s}},\Theta)A^{4}\,.
\end{align}

Given the foregoing generalizations, the perturbation strength dependence of the PIMC data for all the aforementioned energy components should comply to the general form 
\begin{eqnarray}\label{eq:energy_fit}
\Delta{Q}\simeq\gamma_{\mathrm{Q},1}A^2+\gamma_{\mathrm{Q},3}A^4\,,
\end{eqnarray}
where the parameters $\gamma_{\mathrm{Q},1}$ and $\gamma_{\mathrm{Q},3}$ are now treated as free and are obtained from fits to our new PIMC data in the limit of small $A$. The validity of Eq.~(\ref{eq:energy_fit}) is substantiated by the good agreement between fit and data for all considered densities, temperatures, and wave numbers in this work.

\section{Results\label{sec:results}}

Throughout this work, we simulate $N=34$ unpolarized electrons in the canonical ensemble. Finite-size effects in wave-number resolved properties are inconsequential at the conditions considered in this work, which has been investigated in detail in previous works~\cite{dornheim_prl,dornheim_electron_liquid,Dornheim_PRE_2020,Dornheim_PRL_2020}. All the simulation results are freely available online~\cite{online_note} and can be used to benchmark approximations, new methods, and as input for other calculations.

\subsection{Electron liquid regime\label{sec:electron_liquid}}

Let us begin our investigation by discussing PIMC simulations of the strongly coupled electron liquid at a density parameter of $r_s=20$. This regime has attracted considerable interest both in the ground state~\cite{Takada_PRB_2016,Azadi_Wigner_2022,Holzmann_PRL_2020,Spink_PRB_2013} and at finite temperatures~\cite{dornheim_dynamic,Dornheim_Nature_2022,Dornheim_Force_2022,Dornheim_PRR_2022,dornheim_electron_liquid,Tolias_JCP_2021,Tolias_arxiv_2023}, and exhibits interesting features such as the \emph{roton minimum} in the dispersion of density-density fluctuations~\cite{Dornheim_Nature_2022,Takada_PRB_2016,Dornheim_insight_2022,Koskelo}. From a theory perspective, the pronounced effect of electronic correlations in this regime is highly useful to isolate the manifestation of particular features such as the electronic pair alignment, discussed in Ref.~\cite{Dornheim_Nature_2022}, which are also present, though less clearly observable, at metallic densities. Note that our extensive new PIMC data for these states constitute a challenging and unassailable benchmark for the development of new methods and improved dielectric theories~\cite{tanaka_hnc,castello2021classical,Tolias_JCP_2021,Tolias_arxiv_2023}.

\begin{figure}\centering
\hspace*{-0.013\textwidth}\includegraphics[width=0.462\textwidth]{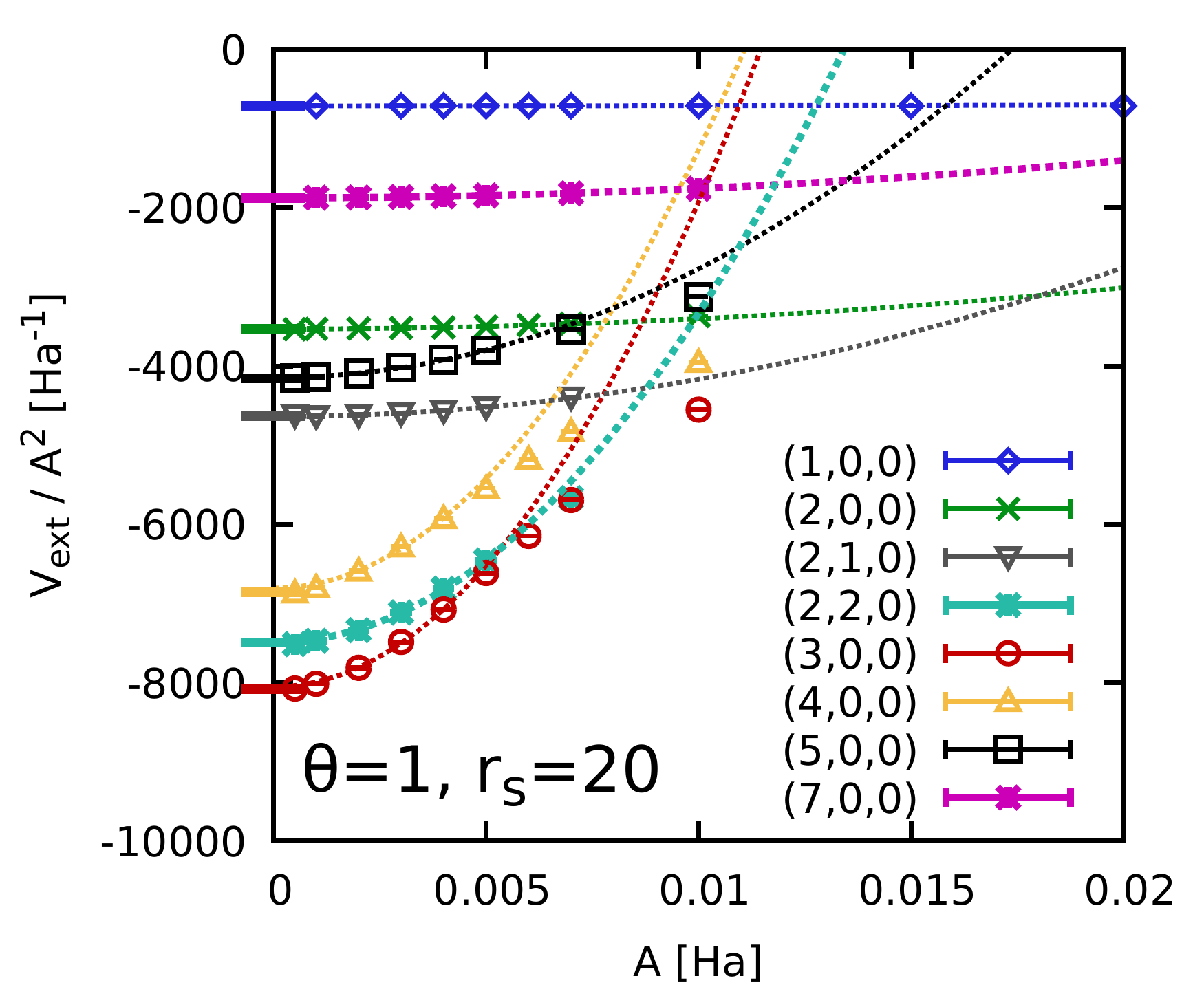}\\\vspace*{-0.9cm}
\includegraphics[width=0.45\textwidth]{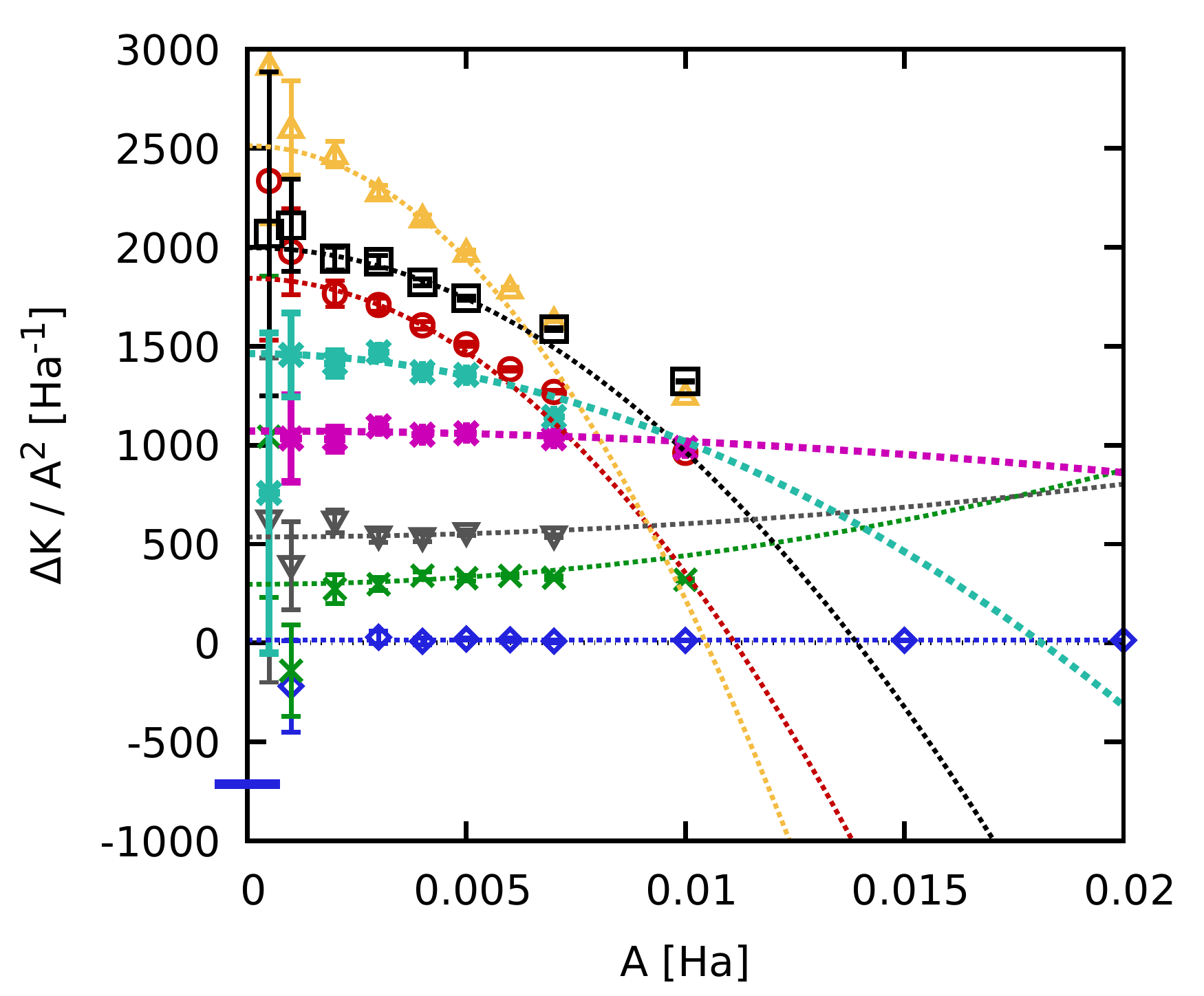}\\\vspace*{-0.9cm}\includegraphics[width=0.45\textwidth]{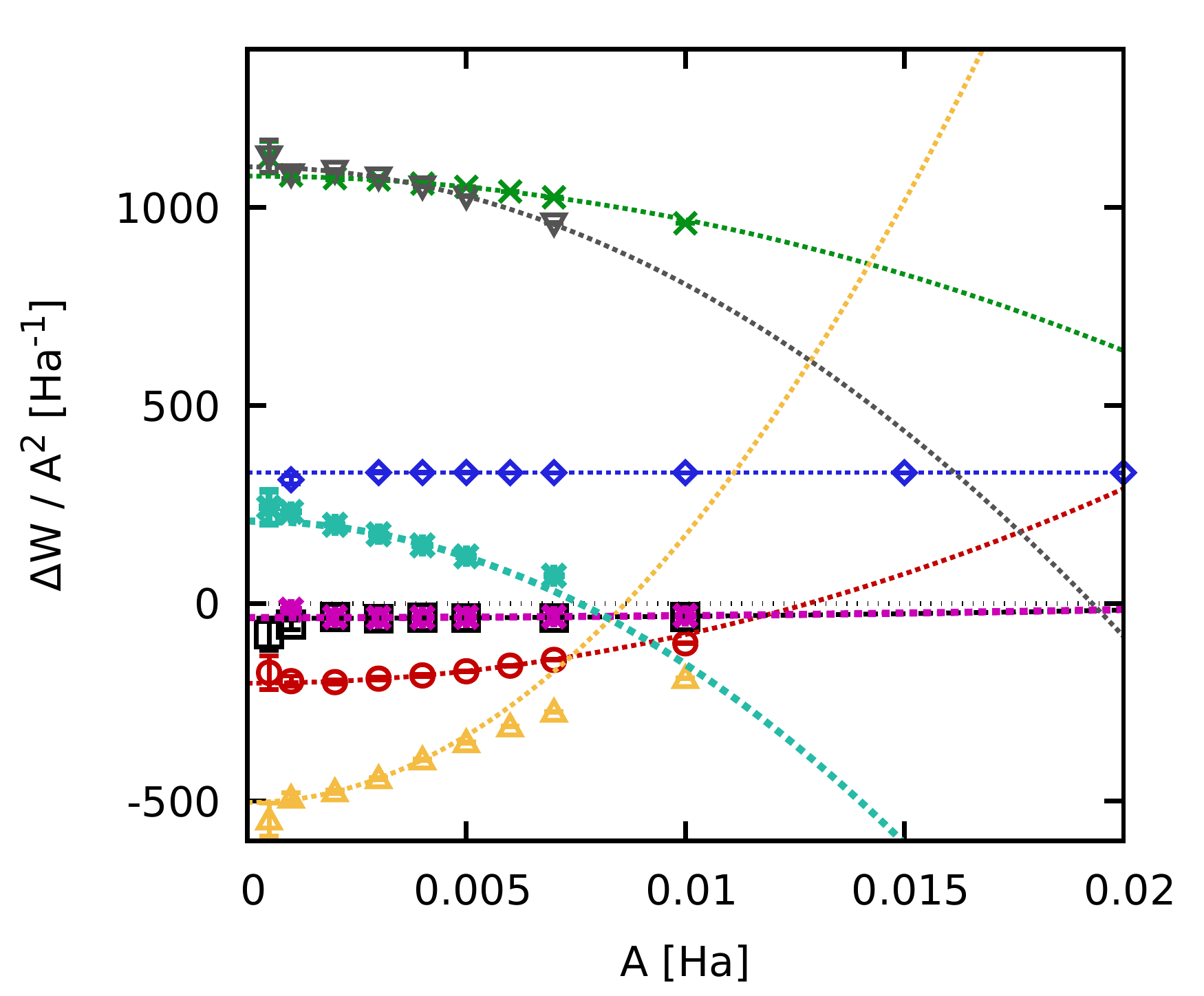}
\caption{\label{fig:N34_rs20_theta1_A}
The induced external potential energy (top), kinetic energy change (center) and interaction energy change (bottom) for $N=34$, $r_s=20$, $\Theta=1$. The symbols depict PIMC data and the dotted curves fits according to Eq.~(\ref{eq:energy_fit}).}
\end{figure}

In Fig.~\ref{fig:N34_rs20_theta1_A}, we show our PIMC simulation results for the dependence of the external potential energy $V_\textnormal{ext}(q,A)$ (top), the kinetic energy change $\Delta K(q,A)$ (center) and the interaction energy change $\Delta W(q,A)$ (bottom) at the electronic Fermi temperature, $\Theta=1$. Specifically, the different symbols/colors correspond to different wave vectors $\mathbf{q}$, and the dotted curves depict fits to these data that are based on Eq.~(\ref{eq:energy_fit}) in a $A\in[0,A_\textnormal{max}]$ interval. The cut-off $A_\textnormal{max}$ has been chosen variably to ensure the absence of significant deviations between fit and data within the given statistical uncertainty. Note that all QMC data have been divided by $A^2$, which means that the curves and data sets approach a constant in the linear response limit, $A\to0$. This theoretical prediction is indeed nicely fulfilled by our PIMC results, as it is expected. Moreover, we stress that each point has been obtained from a single, independent PIMC simulation of the harmonically perturbed UEG governed by the Hamiltonian Eq.~(\ref{eq:Hamiltonian_modified}) with a particular combination of $\mathbf{q}$ and $A$.

Let us first consider the external potential energy $V_\textnormal{ext}$, shown in the top Fig.~\ref{fig:N34_rs20_theta1_A} panel. In the limits of small (blue diamonds) and large wave numbers (purple crosses), the curves are nearly flat, which indicates that nonlinear effects are comparably unimportant over the entire depicted $A$-range. This is consistent to previous studies of the nonlinear density response at metallic densities~\cite{Dornheim_PRL_2020,Dornheim_PRR_2021}. 
The horizontal bars at $A=0$ have been obtained on the basis of PIMC simulations of the unperturbed UEG from the imaginary-time fluctuation--dissipation theorem results for $\chi(\mathbf{q})$ [Eq.~(\ref{eq:static_chi})] into Eq.~(\ref{eq:stiffnessexternal}). Firstly, we find that these independent results are in perfect agreement with the $A\to0$ limit of the present data for the perturbed UEG, which further substantiates the consistency and validity of our set-up. In addition, we find that the linear density response, which directly defines $V_\textnormal{ext}(\mathbf{q},A)$, attains a maximum at intermediate wave numbers. Interestingly, the nonlinear effects do not directly follow the linear response; for example, nonlinear effects are clearly more pronounced for $\mathbf{q}=(5,0,0)^T2\pi/L$ (black squares) compared to $\mathbf{q}=(2,1,0)^T2\pi/L$ (grey downward triangles), even though the linear response is larger in the latter case. A detailed study of nonlinear effects based on the frameworks developed in Refs.~\cite{Dornheim_PRR_2021,Dornheim_JCP_ITCF_2021} is beyond the scope of the present work and will be pursued in dedicated future investigations.

We next consider the induced kinetic energy change, shown in the center Fig.~\ref{fig:N34_rs20_theta1_A} panel. In this case, the changes take place on a smaller energy scale compared to $V_\textnormal{ext}$. Heuristically, this makes intuitive sense as $V_\textnormal{ext}$ is a direct consequence of the external potential, whereas the kinetic energy is only affected indirectly. In addition, our PIMC results are afflicted with a larger statistical uncertainty (both relative and absolute) for this observable. This is a well-known property of the standard thermodynamic estimator of PIMC and can potentially be overcome by a modified virial estimator~\cite{Janke_JCP_1997} in future works. Overall, nonlinear effects for $A>0$ seem to play a similar role compared to $V_\textnormal{ext}$, whereas the linear response is comparably more pronounced at large wave numbers. This is likely a consequence of the single-particle dispersion of the kinetic energy $\omega_\textnormal{K}\sim q^2$.

Finally, the bottom panel of Fig.~\ref{fig:N34_rs20_theta1_A} shows the same information for the interaction energy change $\Delta W(q,A)$. Evidently, changes in $W$ constitute the smallest contribution to the total energy change, although they are on the same scale as $\Delta K(q,A)$ for most depicted wave numbers. Interestingly, $\Delta W$ has a non-monotonic dependence on $q$. More specifically, the interaction energy change due to the external perturbation is positive for small wave numbers, and becomes negative around $\mathbf{q}=(3,0,0)^T2\pi/L$. This is a direct consequence of the pair alignment idea introduced in Ref.~\cite{Dornheim_Nature_2022}, and it is tightly connected to the effective attraction observed in Refs.~\cite{Dornheim_Force_2022,Dornheim_PRR_2022}.

\begin{figure}\centering
\includegraphics[width=0.5\textwidth]{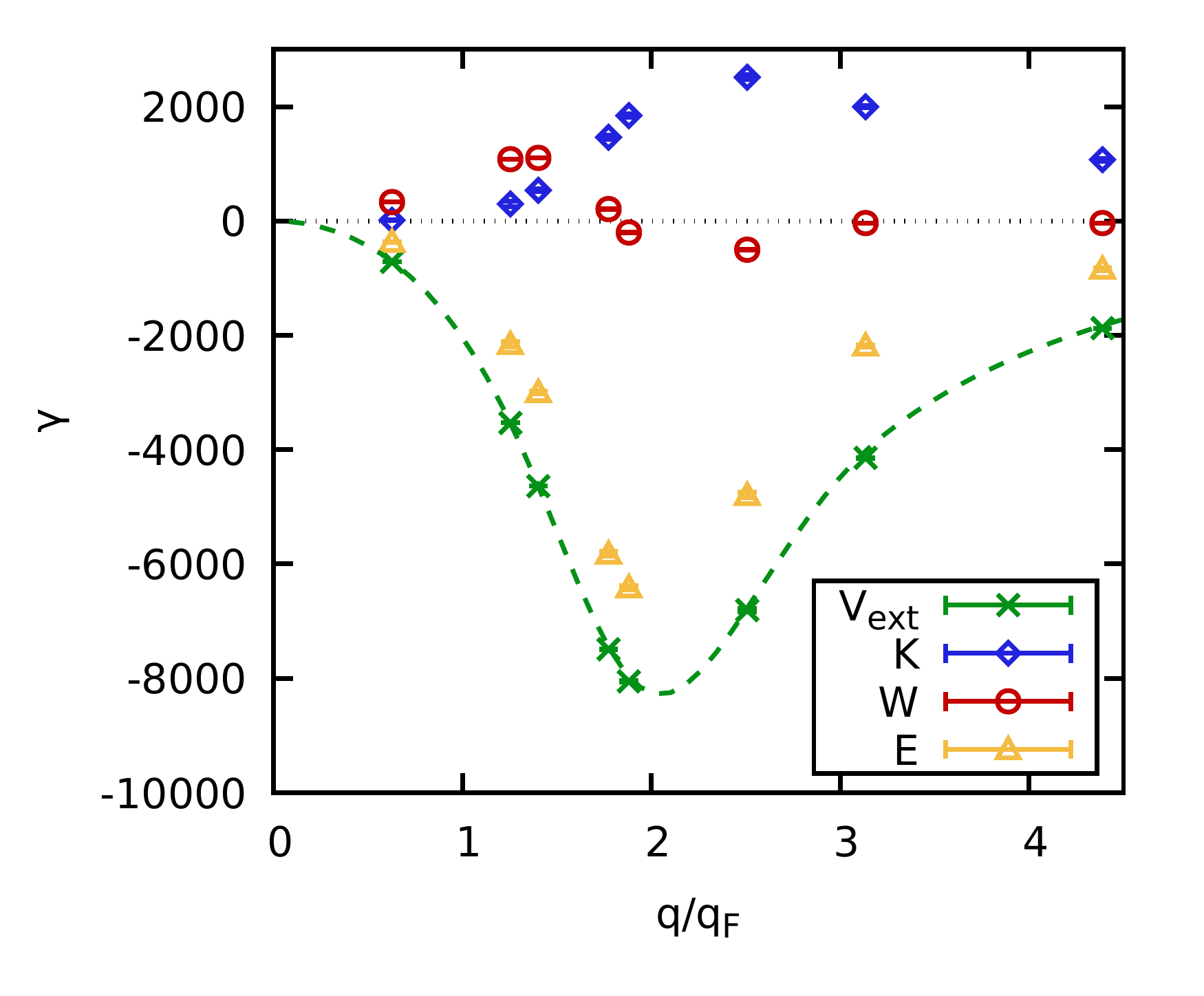}\\\vspace*{-1cm}\hspace*{0.5cm}\includegraphics[width=0.47\textwidth]{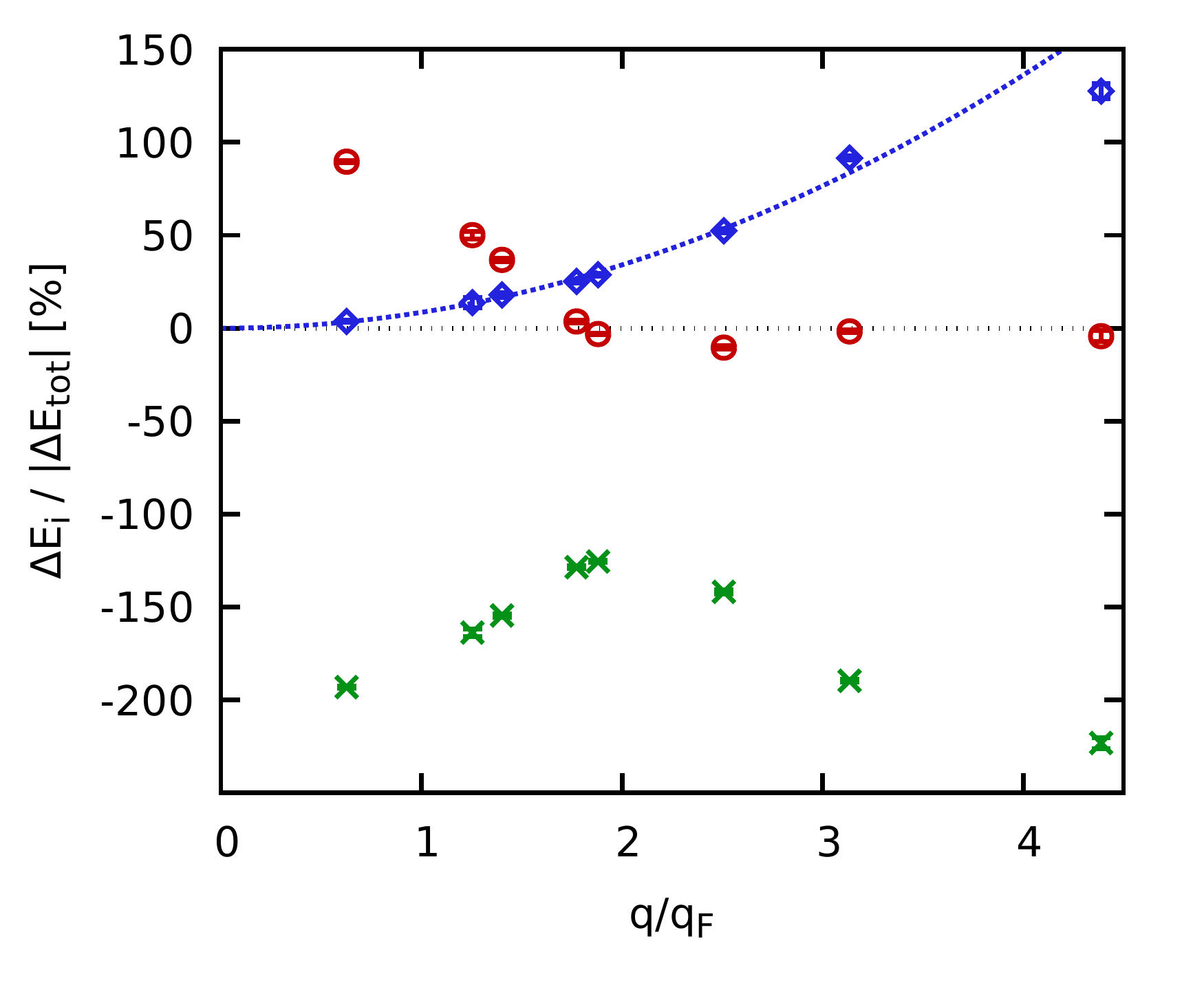}
\caption{\label{fig:N34_rs20_theta1_q}
The wave number dependence of the linear density response limit of different contributions to the induced energy change $\Delta E(q,A)$ for the UEG at $r_s=20$, $\Theta=1$. Top: linear-response coefficients $\gamma_i$ [see Eq.~(\ref{eq:energy_fit})]; bottom: relative contribution of different energies to the change in the total energy. The dotted blue lines depicts a parabolic fit motivated by the single particle dispersion $\omega_K\sim q^2$.}
\end{figure} 

For a more qualitative picture of the Fig.~\ref{fig:N34_rs20_theta1_A} observations, we show the wave number dependence of the linear response coefficients [cf.~Eq.~(\ref{eq:energy_fit}) above] for the interaction energy (red circles), kinetic energy (blue diamonds), and external potential energy (green crosses) in the top panel of Fig.~\ref{fig:N34_rs20_theta1_q}. The latter are in perfect agreement to the corresponding evaluation of the imaginary-time version of the fluctuation--dissipation theorem via Eqs.~(\ref{eq:static_chi}) and (\ref{eq:stiffnessexternal}), as it is expected. It can be clearly discerned that $V_\textnormal{ext}$ constitutes the dominant contribution for all $q$. The kinetic energy change is negligible for small wave numbers $q\lesssim q_\textnormal{F}$ and attains its maximum around $q\sim2.5q_\textnormal{F}$. This is a consequence of two competing effects: 1) the aforementioned single-particle dispersion $\omega_\textnormal{K}\sim q^2$ shifts the maximum in $\Delta K$ to larger $q$, whereas 2) the overall decrease of the density response for $q\gtrsim 2q_\textnormal{F}$ implies that the total $\Delta K$ eventually decreases with $q$.

\begin{figure}\centering
\includegraphics[width=0.5\textwidth]{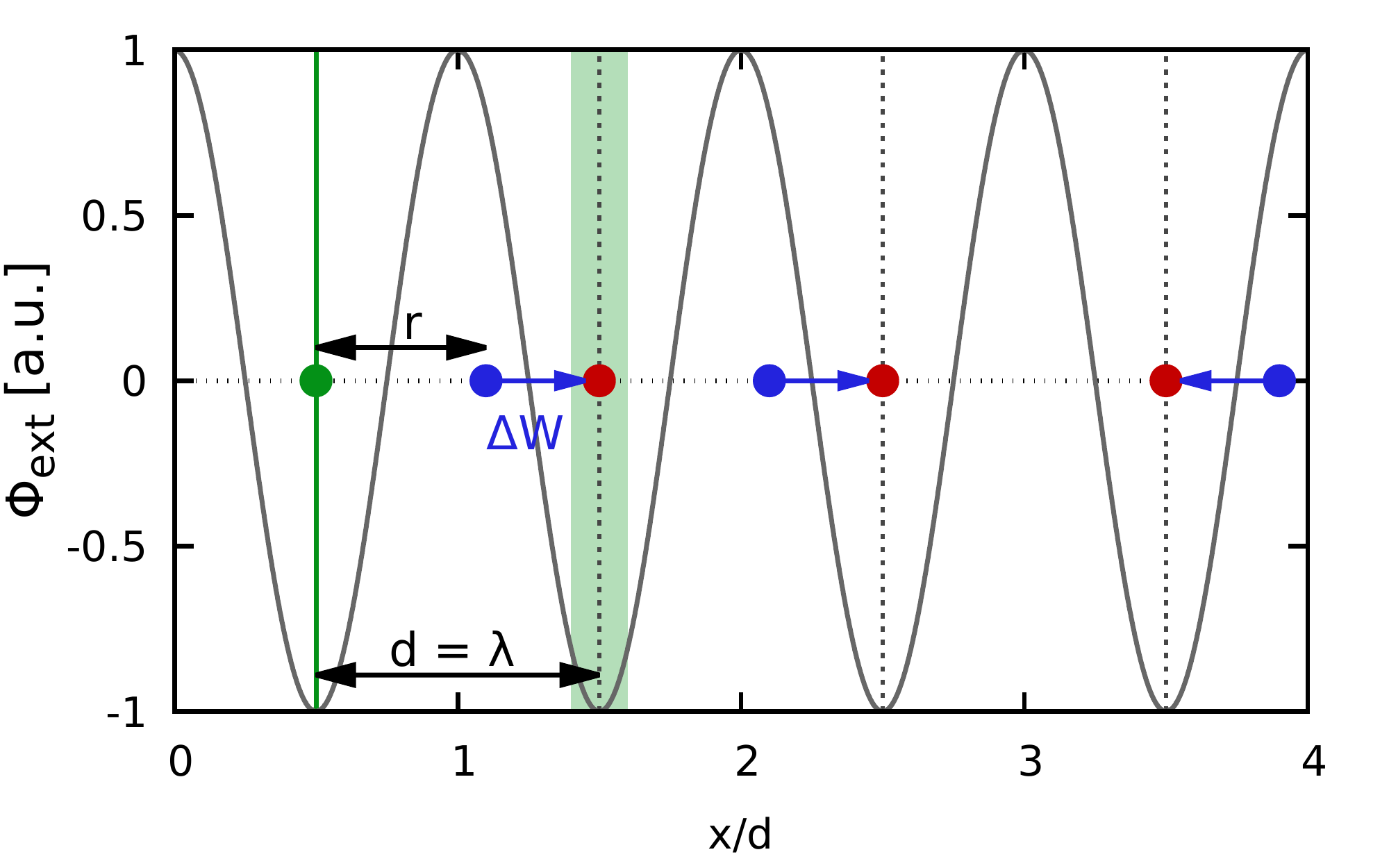}
\caption{\label{fig:sketch}
A schematic illustration of the spatial alignment. In the unperturbed UEG system, the particles are, on average, disordered (blue beads). When the wavelength $\lambda=2\pi/q$ of the external perturbation $\phi_\textnormal{ext}(r)$ (dark grey sinuoidal line) is comparable to the average interparticle separation $d$, the alignment of the particles to the minima of $\phi_\textnormal{ext}$ is associated with a reduction in the interaction energy, $\Delta W<0$.}
\end{figure} 

\begin{figure*}\centering
\includegraphics[width=0.5\textwidth]{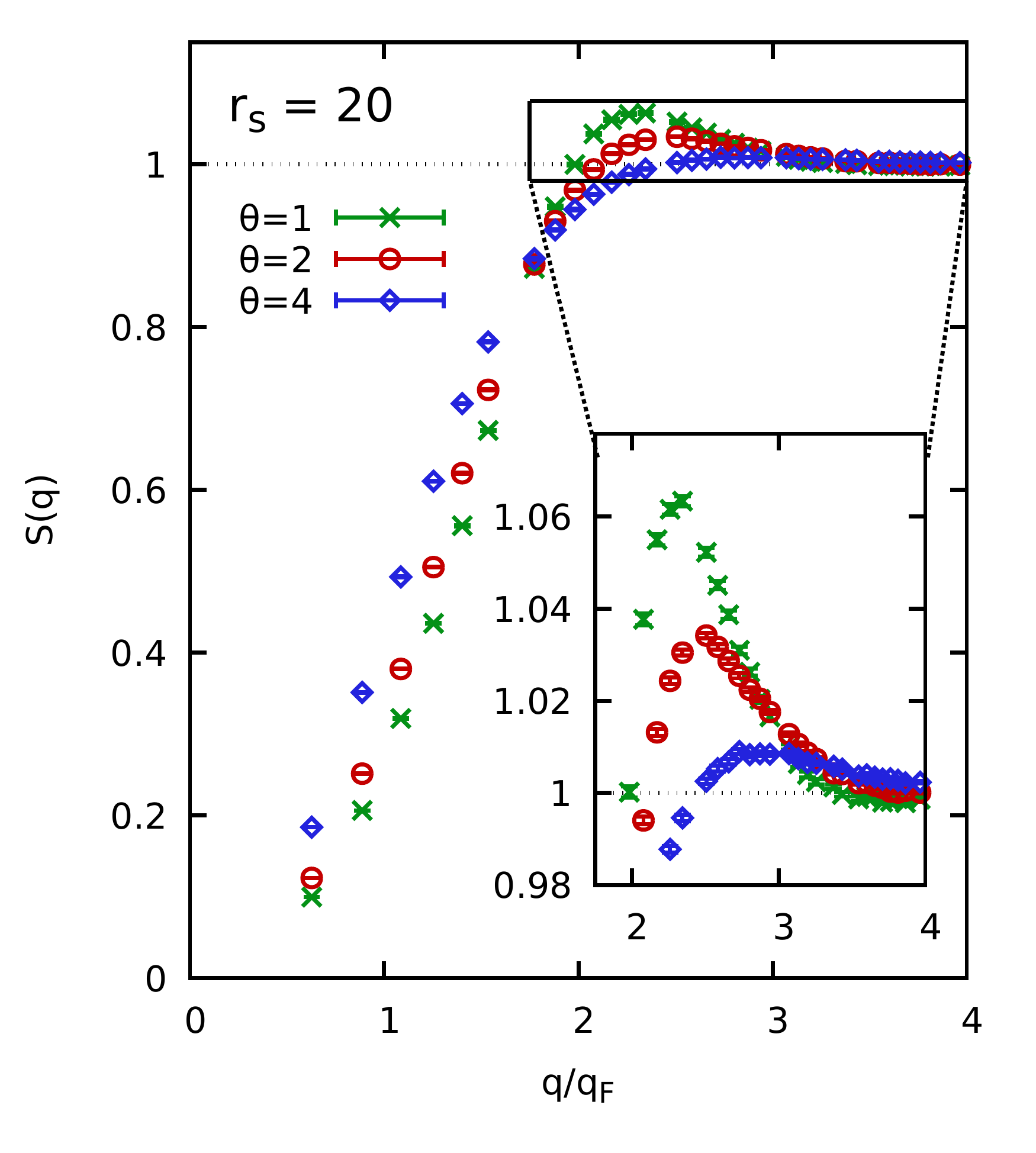}\includegraphics[width=0.5\textwidth]{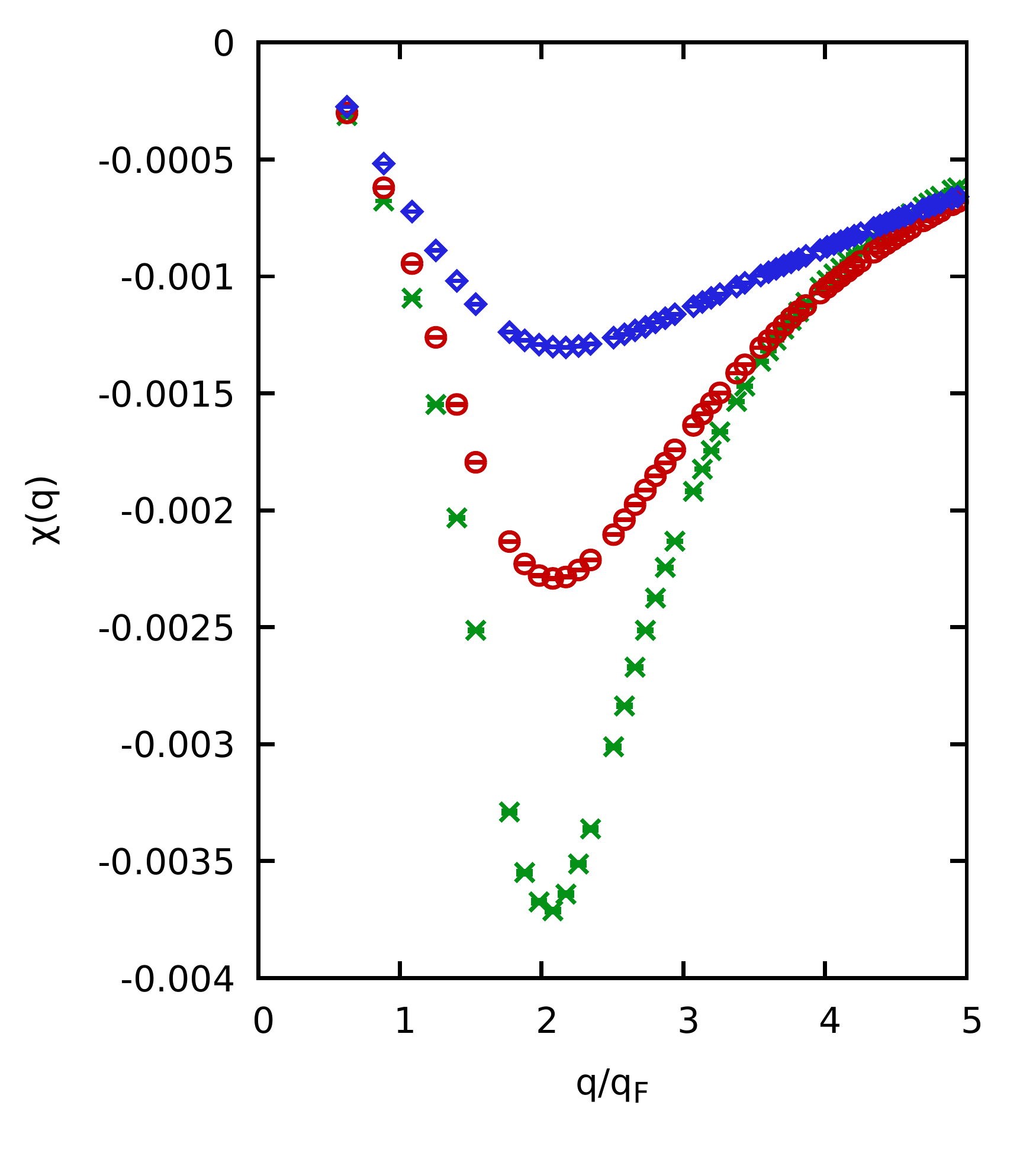}
\caption{\label{fig:N34_rs20_align}
Effect of temperature on the static properties of the unperturbed strongly coupled electron liquid at $r_s=20$. Left: static structure factor $S(\mathbf{q})$, with the inset showing a magnified segment around the peak; right: static linear density response function $\chi(\mathbf{q})$ computed from the imaginary-time density--density correlation function via Eq.~(\ref{eq:static_chi}).}
\end{figure*} 

From a physical perspective, the change in the interaction energy constitutes the most interesting result. In particular, we find a monotonic increase in $\Delta W(q,A)$ with $q$ for $q\lesssim q_\textnormal{F}$, where it attains its maximum. This is followed by a decrease with $q$; indeed, the change in the interaction energy becomes negative for $q\sim2q_\textnormal{F}$. Upon further increasing $q$, the reduction in the interaction energy due to the external harmonic perturbation exhibits a maximum around $q=2.5q_\textnormal{F}$, and then swiftly decreases with $q$. In order to understand these nontrivial observations, in general, and the pronounced reduction in the interaction energy for intermediate wave numbers $q$, in particular, we have to explore the associated length scales. A corresponding schematic illustration is given in Fig.~\ref{fig:sketch}. Without the loss of generality, we select the green bead as an arbitrary reference particle. Without the external harmonic perturbation, the other particles (blue beads) are, on average, disordered at these conditions. This can be discerned from the absence of any pronounced structure from the static structure factor $S(q)$ shown in Fig.~\ref{fig:N34_rs20_align}. If we activate the external potential (dark grey sinusoidal line), the hitherto disordered particles will align themselves to the external potential minima (red beads). For completeness, we note that the depiction of the electrons as classical point particles in Fig.~\ref{fig:sketch} that perfectly align themselves to the minima is a simplification that serves to illustrate the basic underlying mechanism. All quantum delocalization effects are exactly contained in our PIMC simulations. Moreover, the alignment to the $\phi_\textnormal{ext}$ minima is arbitrarily small for $A\to0$.
In the example depicted in Fig.~\ref{fig:sketch}, the external perturbation wave length is comparable to the average interparticle distance, $\lambda\sim d$. In this case, the alignment to the $\phi_\textnormal{ext}$ minima leads to an interaction energy reduction, which explains the negative $\Delta W$ minimum around $q=2.5q_\textnormal{F}$ observed in Fig.~\ref{fig:N34_rs20_theta1_q}.

We conclude the discussion of Fig.~\ref{fig:sketch} by considering the two relevant limiting cases. For $q\gg q_\textnormal{F}$, the wave length of the external potential becomes much smaller than the separation between two particles, $\lambda\ll d$. In this case, the interaction energy is hardly affected by the density response. Consequently, $\Delta W(q,A)$ vanishes much faster with increasing $q$ compared to $\Delta V_\textnormal{ext}(q,A)$ and $\Delta K(q,A)$.
In the limit of $q\lesssim q_\textnormal{F}$, it holds $\lambda \gg d$. In other words, electrons will encounter many neighbours on their way towards the nearest minimum in $\phi_\textnormal{ext}$, which leads to an increase in $\Delta W$. Indeed, this increase in the interaction energy is the basic mechanism behind the perfect screening in the static density of the UEG~\cite{kugler1}.
This can be seen particularly well in the bottom panel of Fig.~\ref{fig:N34_rs20_theta1_q}, where we show the linear-response limit of the induced changes in the different contributions to the energy normalized by the change in the total energy $\Delta E(q,A)$. This re-scaling clearly reveals the increasing importance of the interaction energy in the limit of $q\to 0$, which, in fact, leads to a vanishing density response in the long wave-length limit. The relative importance of the kinetic energy, on the other hand, monotonically increases with $q$ and nicely follows a parabolic fit motivated by the simple quadratic single-particle dispersion $\omega_K\sim q^2$.

\subsection{Effect of the temperature\label{sec:temperature}}

Let us next investigate the dependence of the induced energy change and the spatial alignment on the temperature. In Fig.~\ref{fig:N34_rs20_align}, we provide PIMC results for the static structure factor $S(\mathbf{q})$ (left) and the linear static density response function $\chi(\mathbf{q})$ of the unperturbed UEG [evaluated from Eq.~(\ref{eq:static_chi})] (right) at $r_s=20$ for $\Theta=4$ (blue diamonds), $\Theta=2$ (red circles), and $\Theta=1$ (green crosses). As mentioned above, the static structure factor does not exhibit any pronounced exchange--correlation features at these conditions, and its peak around $q=2.3q_\textnormal{F}$ does not exceed $1.1$ even for the lowest temperature. This further substantiates our earlier interpretation about the reduction in the interaction energy due to an ordering induced by the external harmonic perturbation. The effect of the temperature is more pronounced on the static density response $\chi(\mathbf{q})$. In particular, the pronounced minimum of $\chi(\mathbf{q})$ is reduced by more than a factor of three between $\Theta=1$ and $\Theta=4$. In contrast, temperature effects nearly vanish in the limits of small and large $q$.

\begin{figure}\centering
\includegraphics[width=0.5\textwidth]{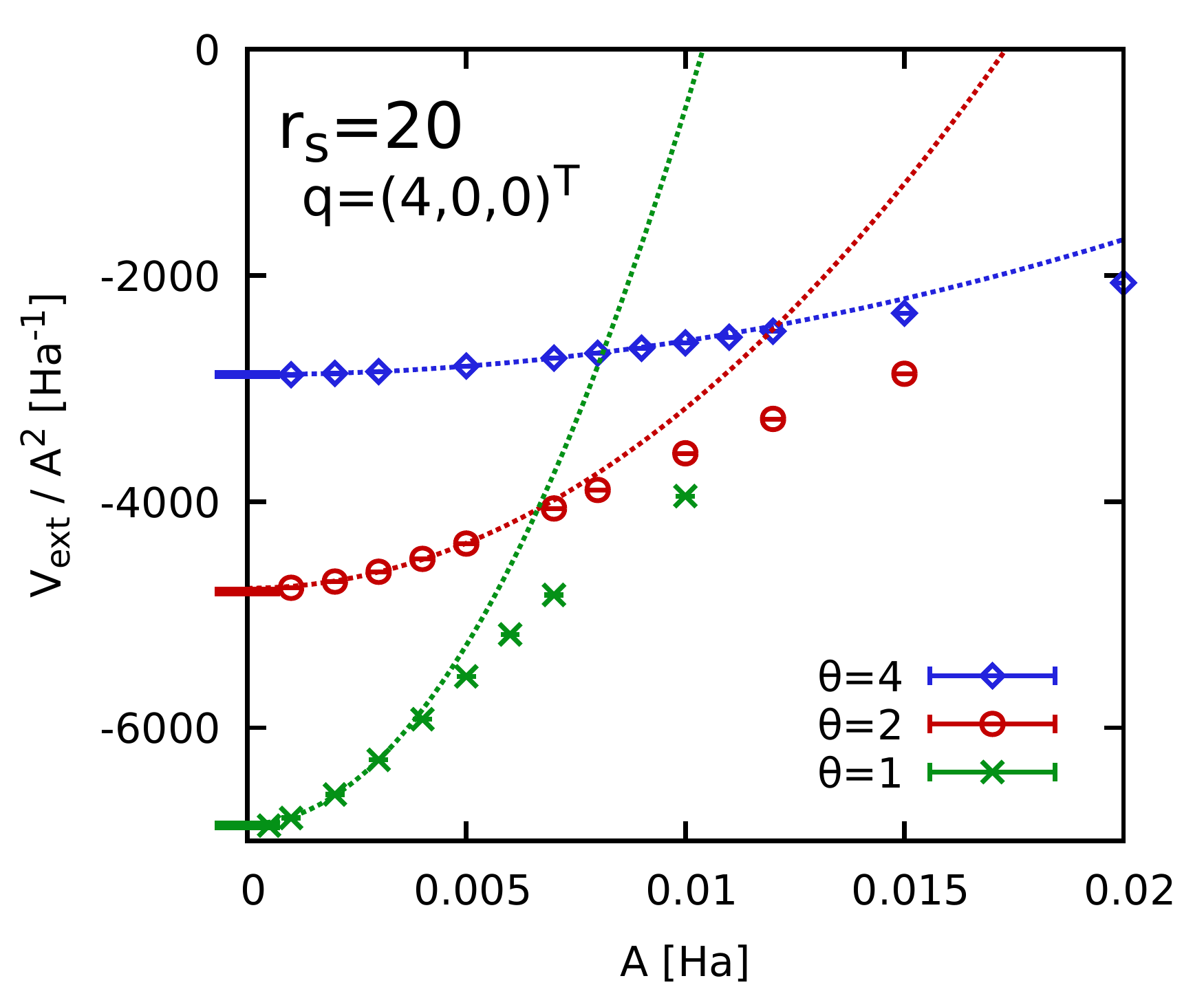}\\\vspace*{-1cm}\includegraphics[width=0.5\textwidth]{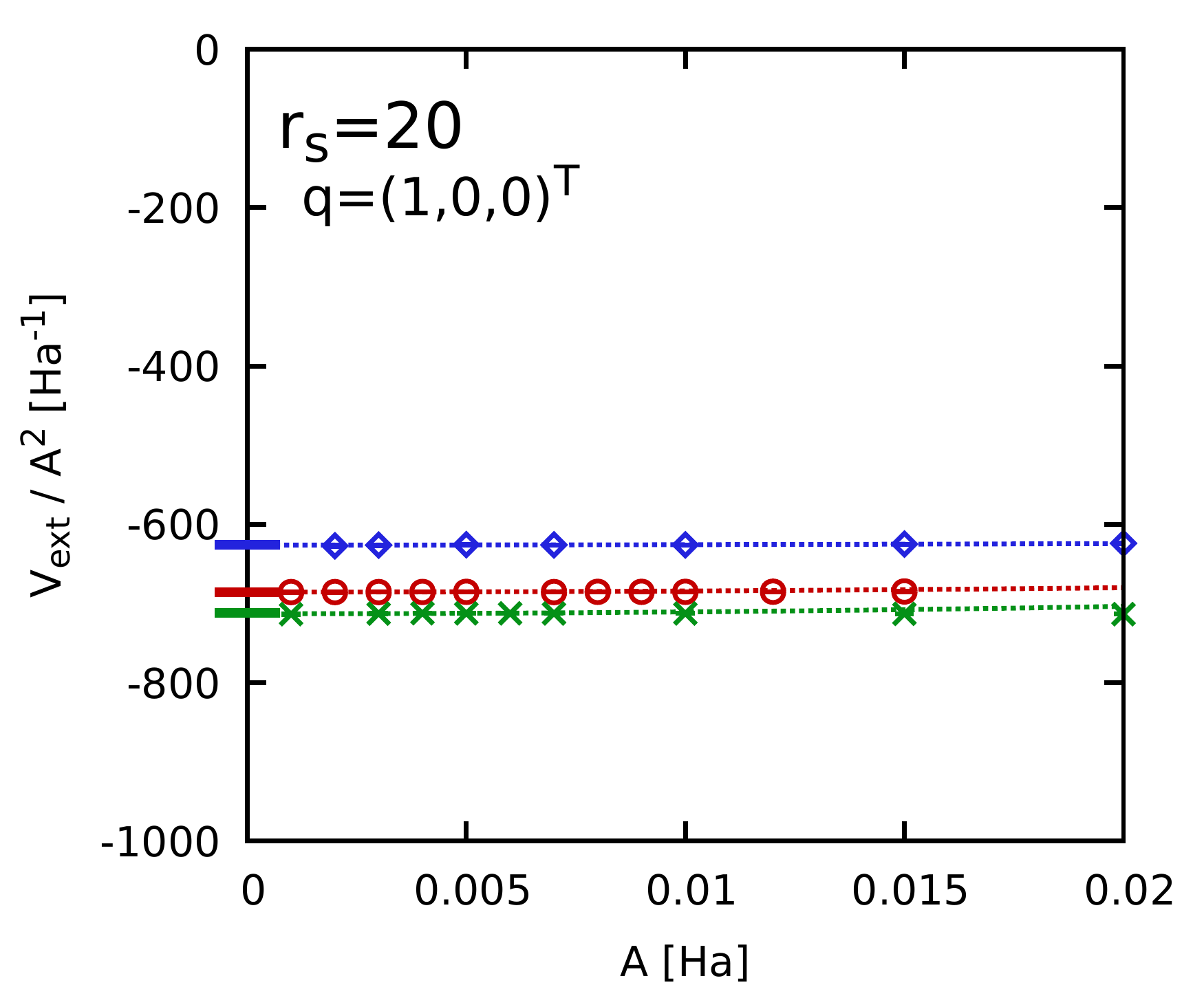}
\caption{\label{fig:Theta_Vext}
Effect of temperature on the external potential energy at $q=2.54q_\textnormal{F}$ and $q=0.63q_\textnormal{F}$ for $r_s=20$. The symbols show raw PIMC simulation data, and the dotted curves analytical fits according to Eq.~(\ref{eq:energy_fit}).}
\end{figure} 

These trends can be identified in our new PIMC results for the $A$-dependence of $V_\textnormal{ext}$, which are shown in Fig.~\ref{fig:Theta_Vext} for $\mathbf{q}=(4,0,0)^T2\pi/L$ (top) and $\mathbf{q}=(1,0,0)^T2\pi/L$ (bottom). In particular, we find a pronounced dependence on $\Theta$ both of the linear-response limit and also the nonlinear contributions for $q=2.54q_\textnormal{F}$, whereas the results for $\Theta=1$ and $\Theta=2$ are almost identical for $q=0.63q_\textnormal{F}$. Moreover, nonlinear effects are nearly absent in the latter regime as the cubic density response function at the first harmonic is substantially more peaked compared to $\chi(\mathbf{q})$ and nearly zero for small and large wave numbers~\cite{Dornheim_PRL_2020,Dornheim_PRR_2021}.

\begin{figure}\centering
\includegraphics[width=0.5\textwidth]{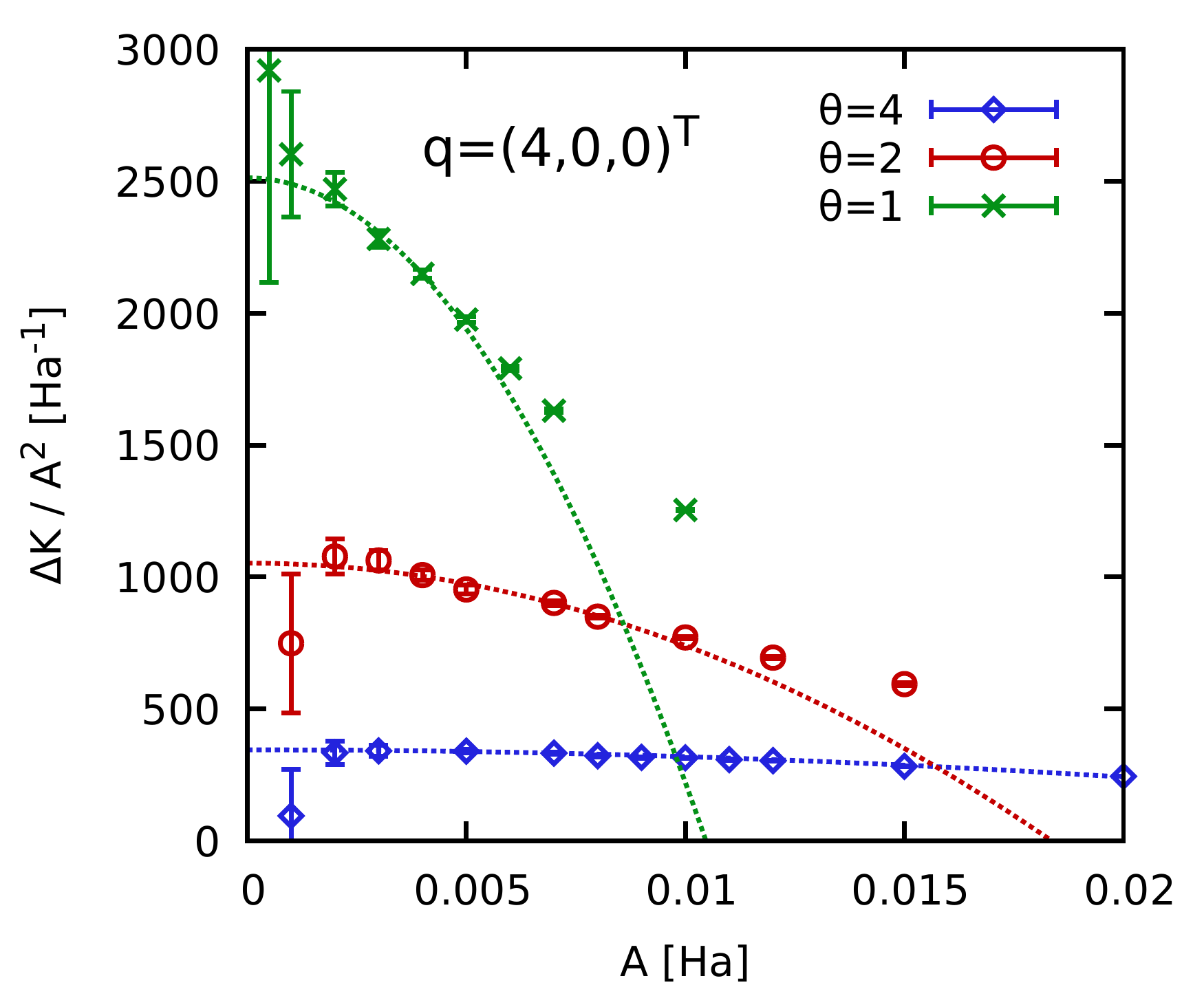}\\\vspace*{-1cm}\hspace*{0.028\textwidth}\includegraphics[width=0.47\textwidth]{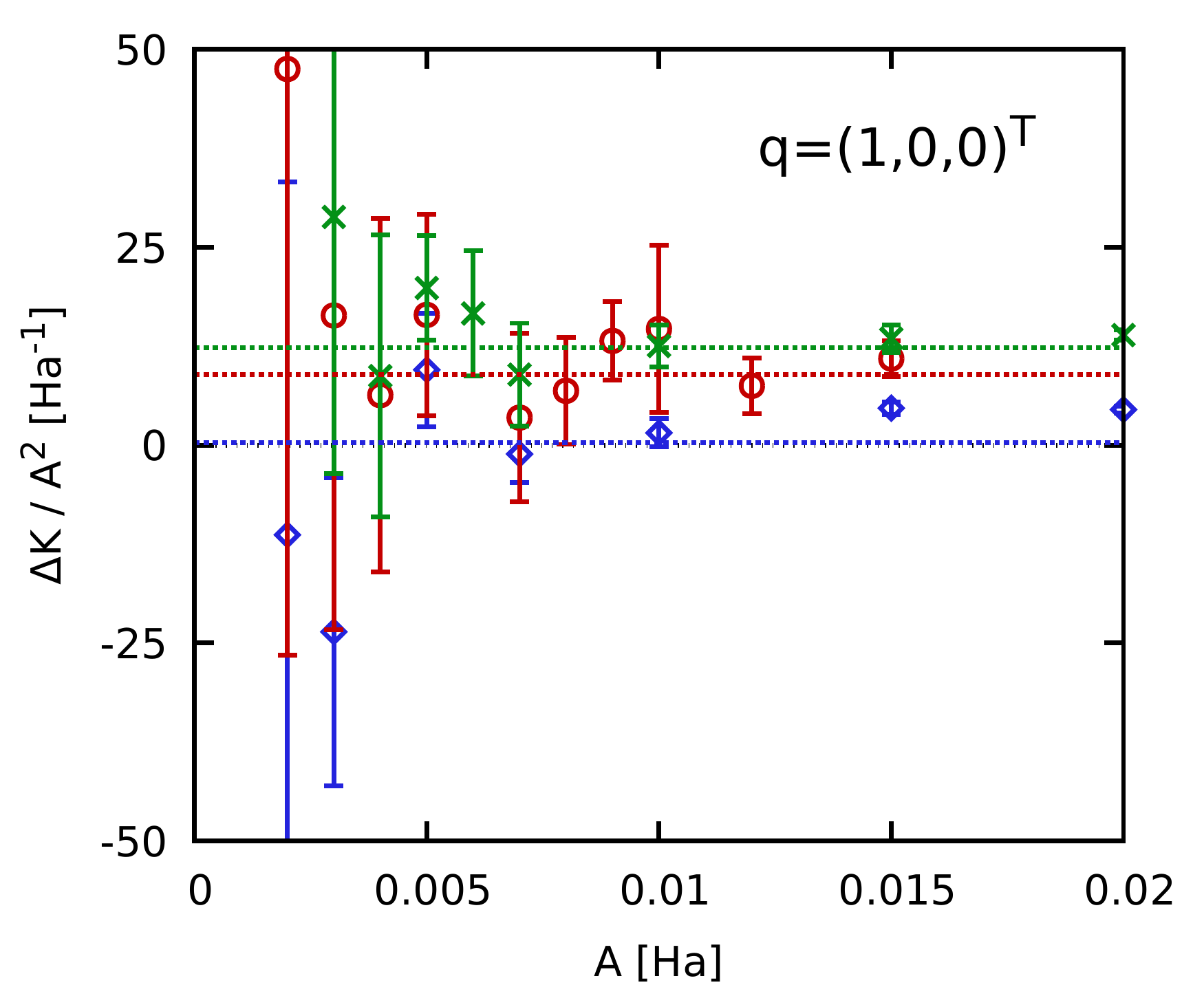}
\caption{\label{fig:Theta_K}
Effect of temperature on the induced kinetic energy change at $q=2.54q_\textnormal{F}$ and $q=0.63q_\textnormal{F}$ for $r_s=20$. The symbols show raw PIMC simulation data, and the dotted curves analytical fits according to Eq.~(\ref{eq:energy_fit}).}
\end{figure} 

In Fig.~\ref{fig:Theta_K}, we repeat this analysis for the kinetic energy change $\Delta K(q,A)$ at the same wave numbers. Overall, the trends are similar to our previous observations for $V_\textnormal{ext}(q,A)$. Specifically, we find the largest change in the kinetic energy, and also by far the largest impact of nonlinear effects at the larger wave number for $\Theta=1$. For $\mathbf{q}=(1,0,0)^T2\pi/L$, the change in the kinetic energy can hardly be resolved, in particular for $\Theta=4$. We note that application of an external perturbation always increases the kinetic energy independent of the temperature.

\begin{figure}\centering
\includegraphics[width=0.5\textwidth]{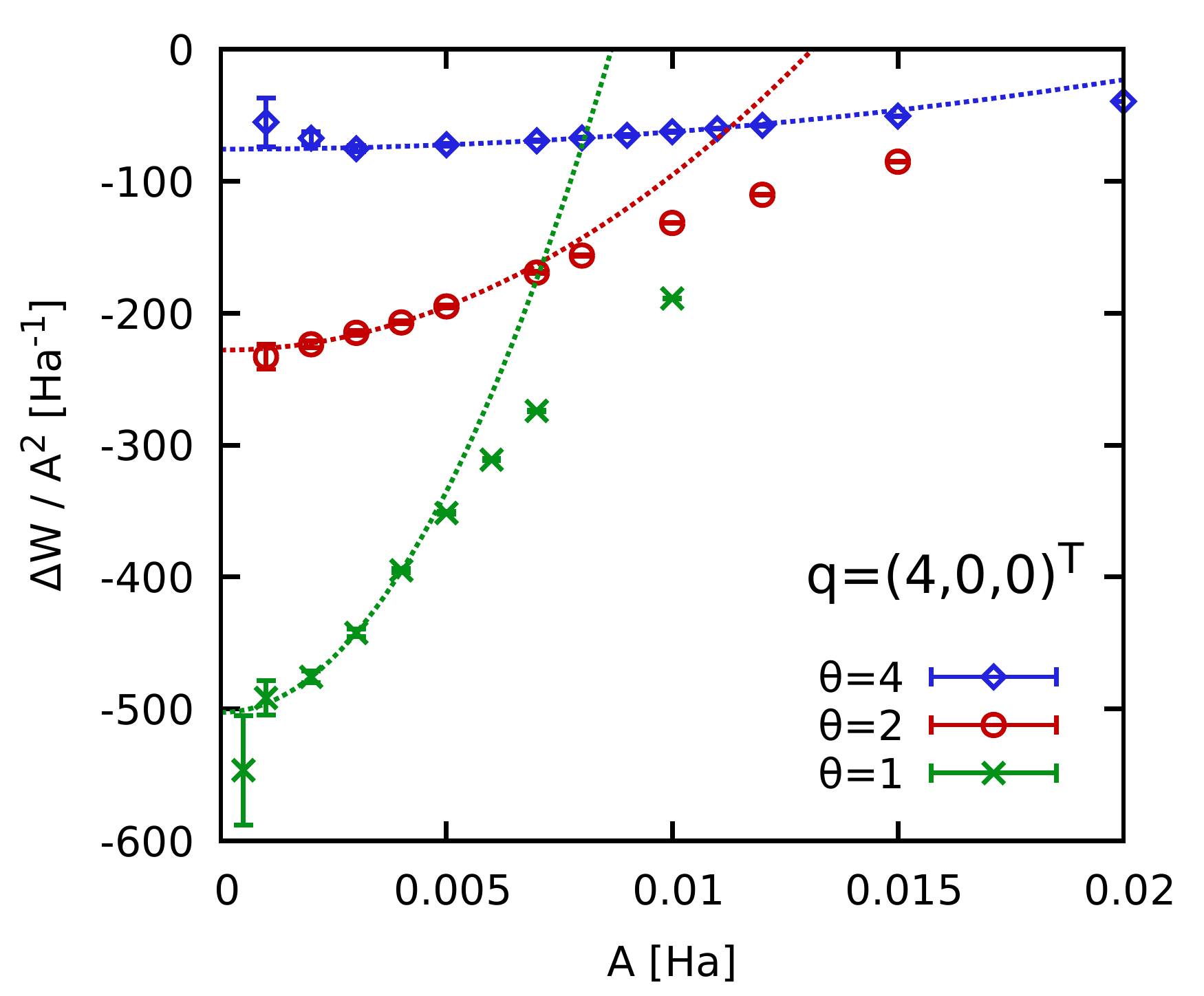}\\\vspace*{-1cm}\hspace*{0.003\textwidth}\includegraphics[width=0.495\textwidth]{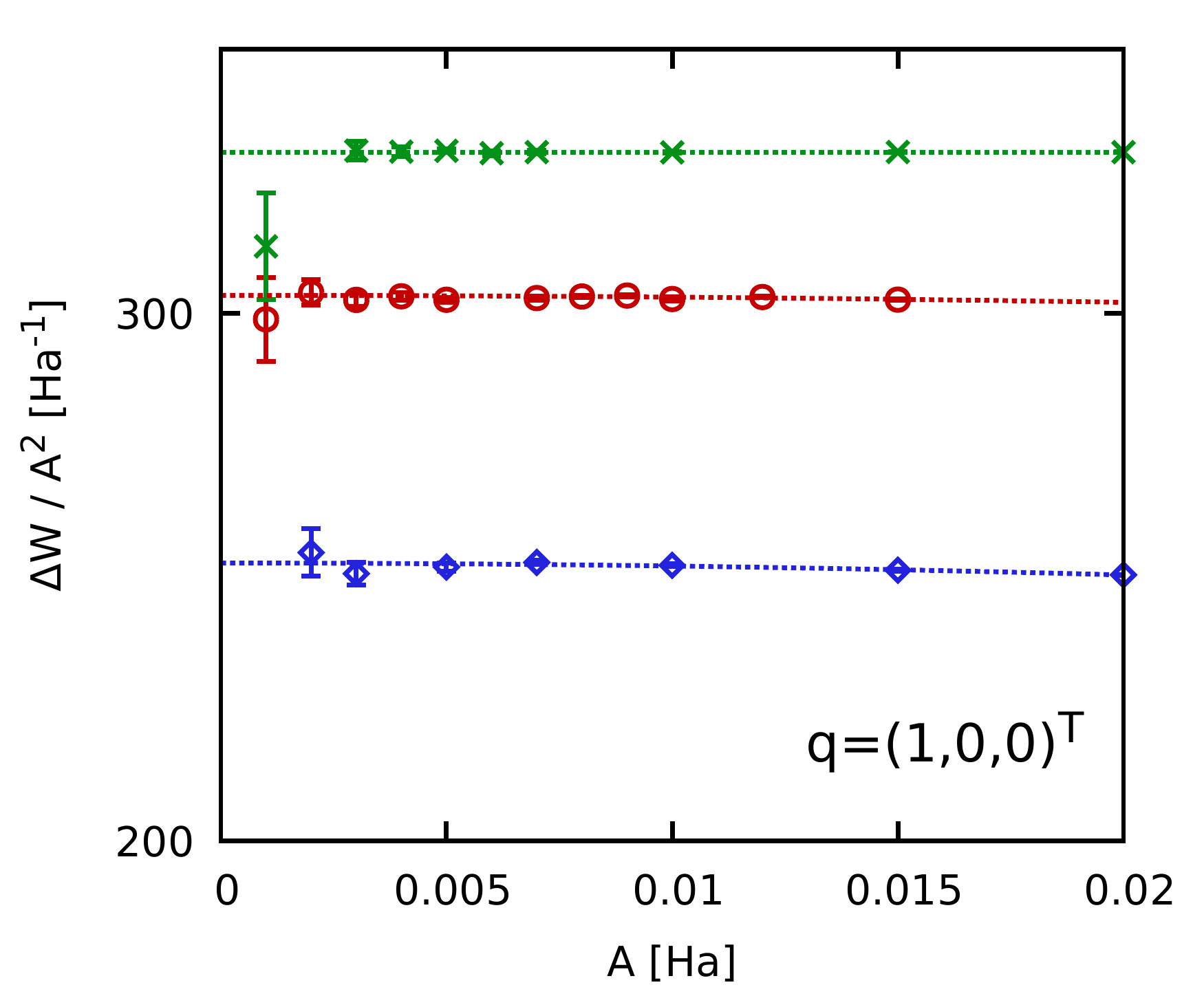}
\caption{\label{fig:Theta_W}
Effect of temperature on the induced interaction energy change at $q=2.54q_\textnormal{F}$ \& $q=0.63q_\textnormal{F}$ for $r_s=20$. The symbols show raw PIMC simulation data, and the dotted curves analytical fits according to Eq.~(\ref{eq:energy_fit}).}
\end{figure} 

Finally, in Fig.~\ref{fig:Theta_W}, we investigate the effect of the temperature on the interaction energy change $\Delta W(q,A)$. For $q=2.54q_\textnormal{F}$, the interaction energy change is negative for all temperatures, although the exchange--correlation induced reduction due to spatial alignment is substantially reduced for $\Theta=4$. In contrast, the change is positive for $q=0.63q_\textnormal{F}$, and the effect of the temperature is significantly less pronounced. From a physical perspective, this can be explained by the different physical mechanisms operative in these two wave-number regimes. For $\lambda\sim d$, the response is strongly affected by correlations, which become reduced with increasing temperature. For $q<q_\textnormal{F}$, the main mechanism is the perfect screening in the UEG~\cite{kugler_bounds}, which depends less sensitively on $\Theta$.

\begin{figure}\centering
\includegraphics[width=0.5\textwidth]{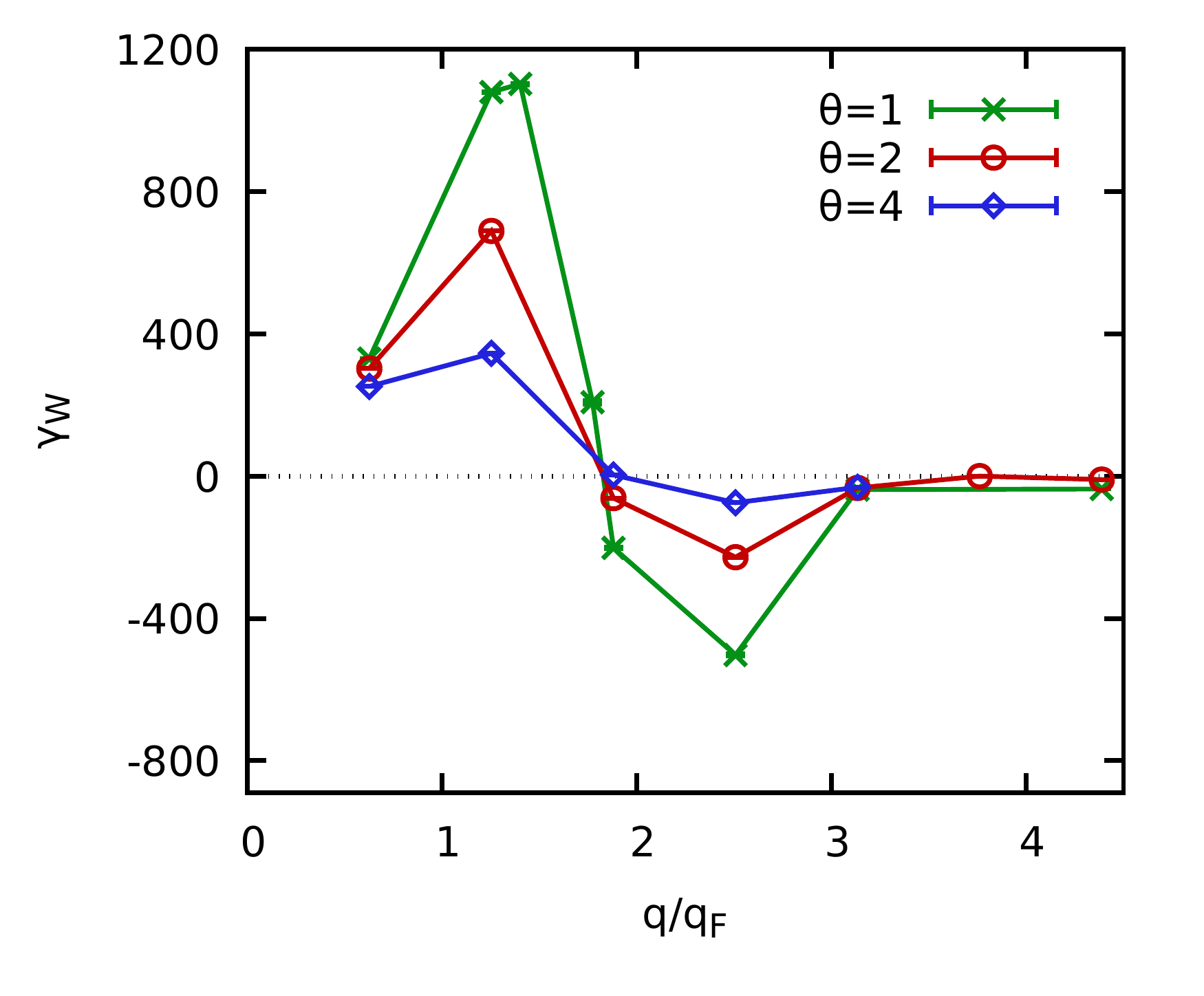}\\\vspace*{-1.31cm}\hspace*{0.003\textwidth}\includegraphics[width=0.495\textwidth]{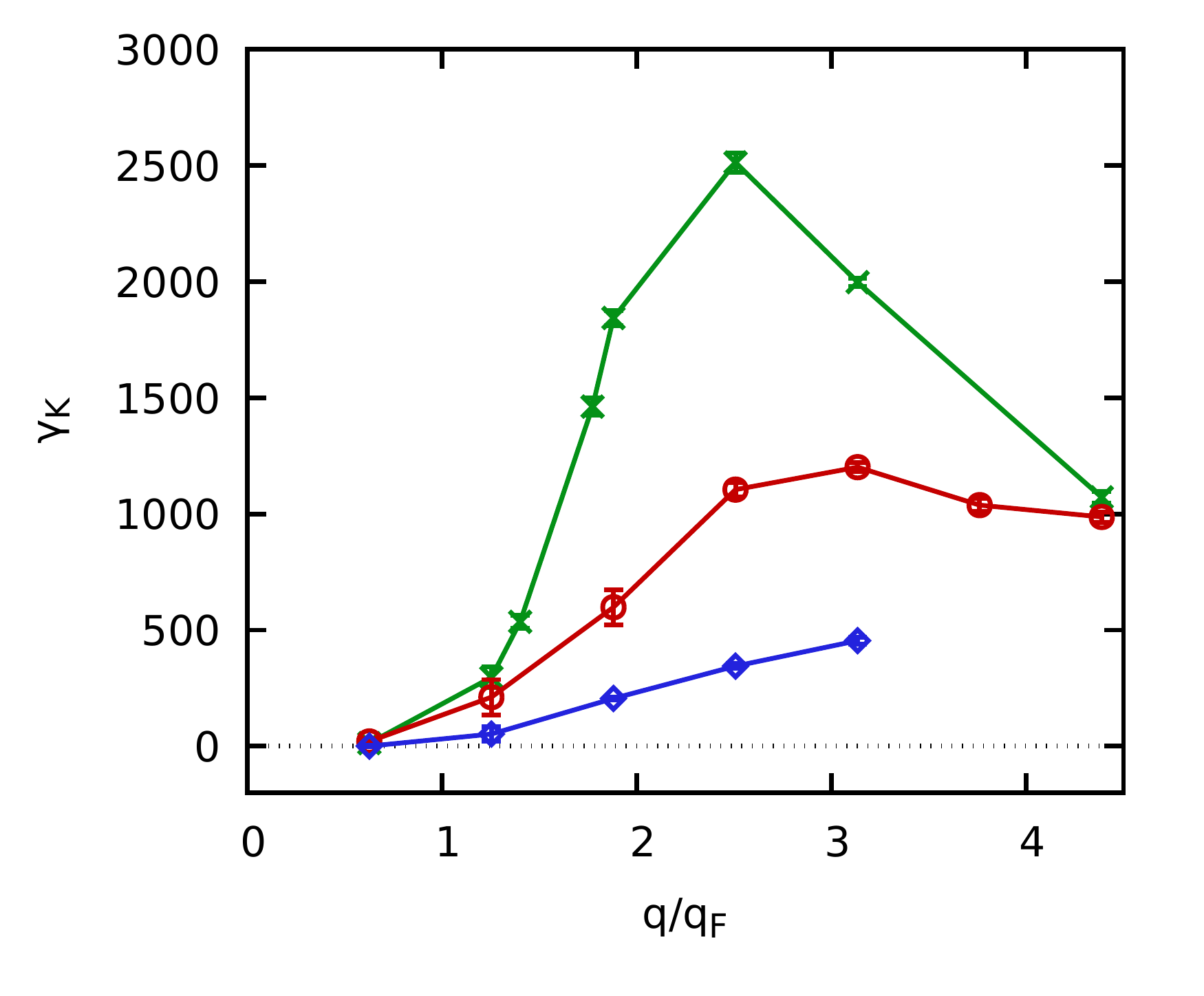}
\caption{\label{fig:Theta_FIT}
Effect of temperature on the linear response coefficients $\gamma_i$ [see Eq.~(\ref{eq:energy_fit})] of the interaction energy (top) and kinetic energy change (bottom) for $r_s=20$.}
\end{figure} 

To get a more qualitative insight into the impact of the temperature on the energy response at different wave numbers, we show the linear-response coefficient $\gamma_i$ [see Eq.~(\ref{eq:energy_fit}) above] of the interaction energy change (top) and the kinetic energy change (bottom) in Fig.~\ref{fig:Theta_FIT}. A corresponding analysis of the external potential energy $V_\textnormal{ext}$ would be redundant, as it exactly follows the static linear density response function $\chi(\mathbf{q})$ shown in Fig.~\ref{fig:N34_rs20_align}. First, we find a pronounced negative minimum in $\Delta W$ around $q=2.5q_\textnormal{F}$ due to the spatial alignment for all three values of $\Theta$, and the depth of the minimum strongly depends on the temperature. The effect vanishes around $q=3q_\textnormal{F}$ for all $\Theta$. Similarly, the interaction energy is increased by the external perturbation for $q\lesssim2q_\textnormal{F}$ for all $\Theta$. Overall, the temperature impact on the kinetic energy is less pronounced for intermediate wave numbers compared to the interaction energy, but still significant.

\subsection{Effect of the density\label{sec:coupling}}

\begin{figure*}\centering
\includegraphics[width=0.5\textwidth]{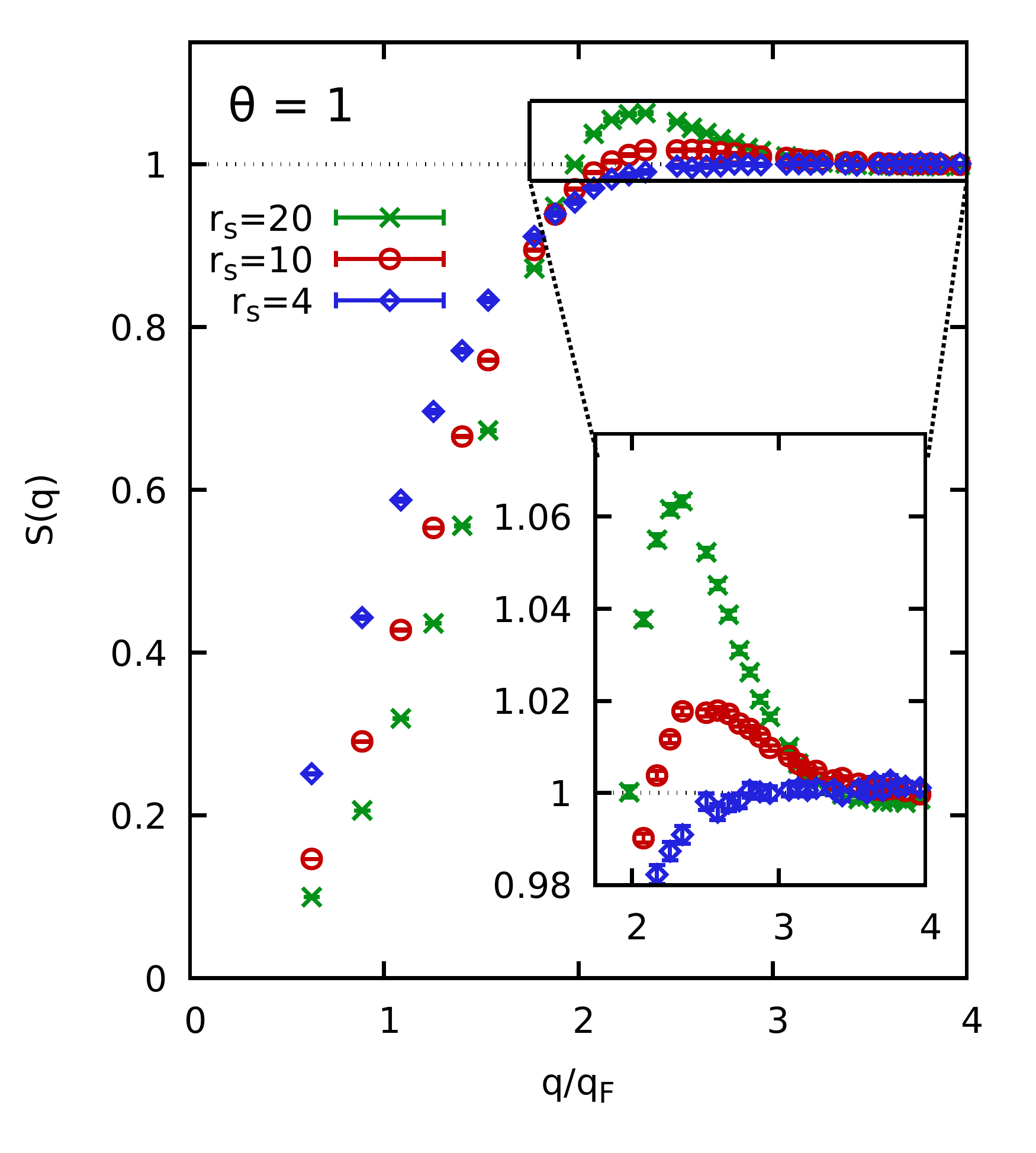}\includegraphics[width=0.5\textwidth]{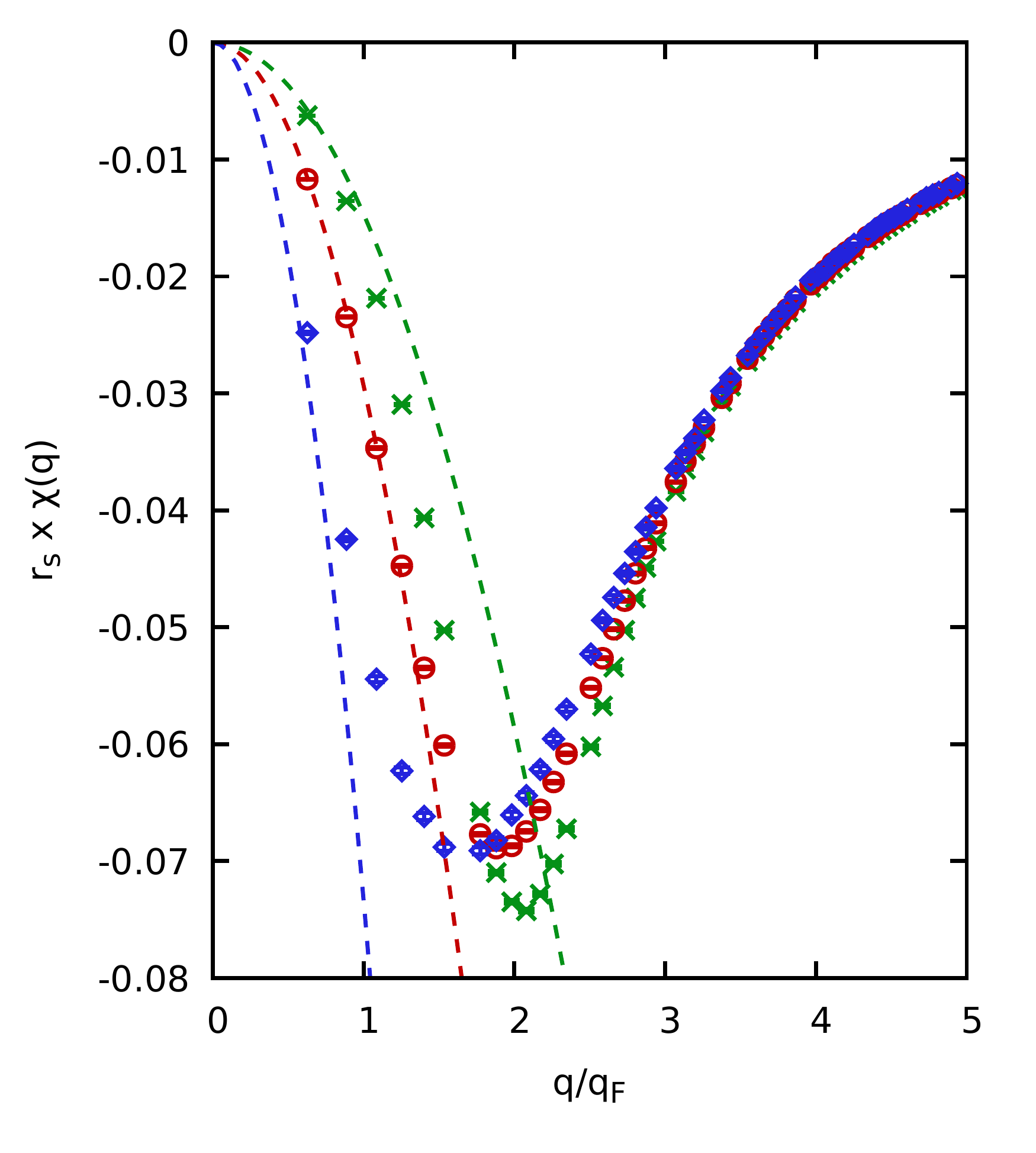}
\caption{\label{fig:N34_theta1_align}
Density dependence of static properties of the unperturbed UEG at $\Theta=1$. Left: static structure factor $S(\mathbf{q})$; right: static linear density response function $\chi(\mathbf{q})$, computed from Eq.~(\ref{eq:static_chi}). The dashed lines correspond to the exact small-$q$ limit of $\chi(\mathbf{q})$ given by Eq.~(\ref{eq:chi_parabola}).}
\end{figure*} 

A further important research question concerns the dependence of the energy response and the corresponding spatial alignment on the density parameter $r_s$. Let us start this investigation by re-calling the behaviour of the familiar static structure factor $S(\mathbf{q})$ and the static linear density response function $\chi(\mathbf{q})$, shown in Fig.~\ref{fig:N34_theta1_align}. In particular, the green crosses correspond to the strongly coupled case of $r_s=20$, the red circles to $r_s=10$, which can be considered as the boundary of the electron liquid regime~\cite{dornheim_dynamic}, and the blue diamonds to the metallic density of $r_s=4$, which is close to sodium~\cite{Huotari_PRL_2010,felde}. With increasing density, the $S(\mathbf{q})$ peak around $q=2.5q_\textnormal{F}$ is substantially decreased at $r_s=10$, and is entirely absent within the given Monte Carlo error bars at $r_s=4$; this is a direct consequence of the role of $r_s$ as the quantum coupling parameter of the UEG.
To allow for a direct comparison of the static linear density response function $\chi(\mathbf{q})$ (right panel), we multiply the latter by a factor of $r_s$; this would lead to identical curves for the ideal (i.e., Lindhard) response function $\chi_0(\mathbf{q})$ for the same $\Theta$ value. First, we note that all curves approach the same noninteracting limit for $q\gg q_\textnormal{F}$. Second, all curves exhibit a negative minimum around $q=2q_\textnormal{F}$ of similar magnitude, and the position is slightly shifted towards smaller wave numbers with increasing density, i.e., decreasing $r_s$. Finally, we observe that, while the static density response always vanishes for $q=0$ due to the perfect screening in the UEG, this happens more rapidly for lower densities. This is a direct consequence of dividing the wave number by $q_\textnormal{F}$ in Fig.~\ref{fig:N34_theta1_align}, as all curves attain the same small-$q$ limit (dashed lines) given by~\cite{kugler_bounds}
\begin{eqnarray}\label{eq:chi_parabola}
\lim_{q\to0}\chi(\mathbf{q}) = -\frac{q^2}{4\pi}\ .
\end{eqnarray}

\begin{figure}\centering
\includegraphics[width=0.5\textwidth]{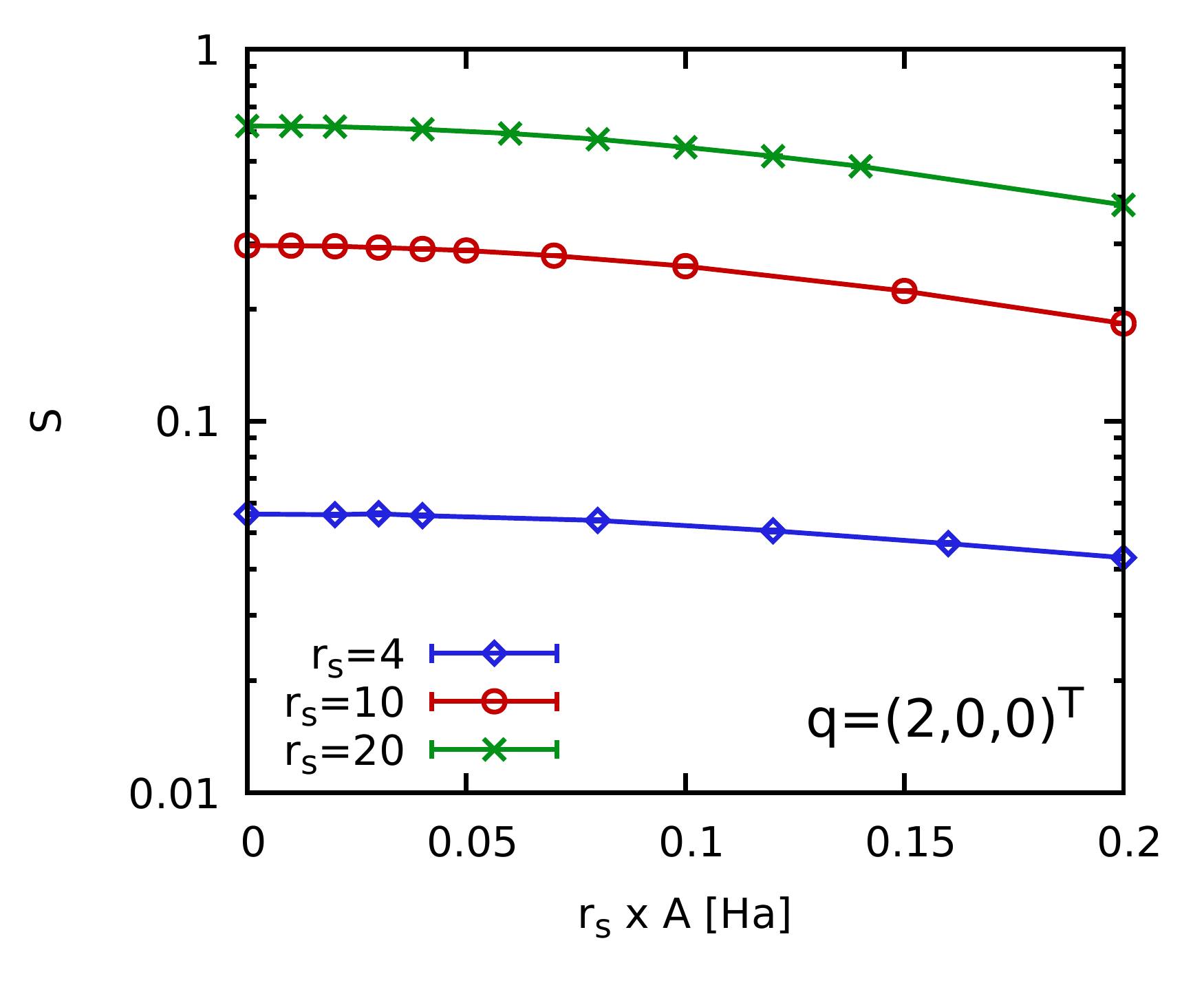}\\\vspace*{-1.31cm}\includegraphics[width=0.5\textwidth]{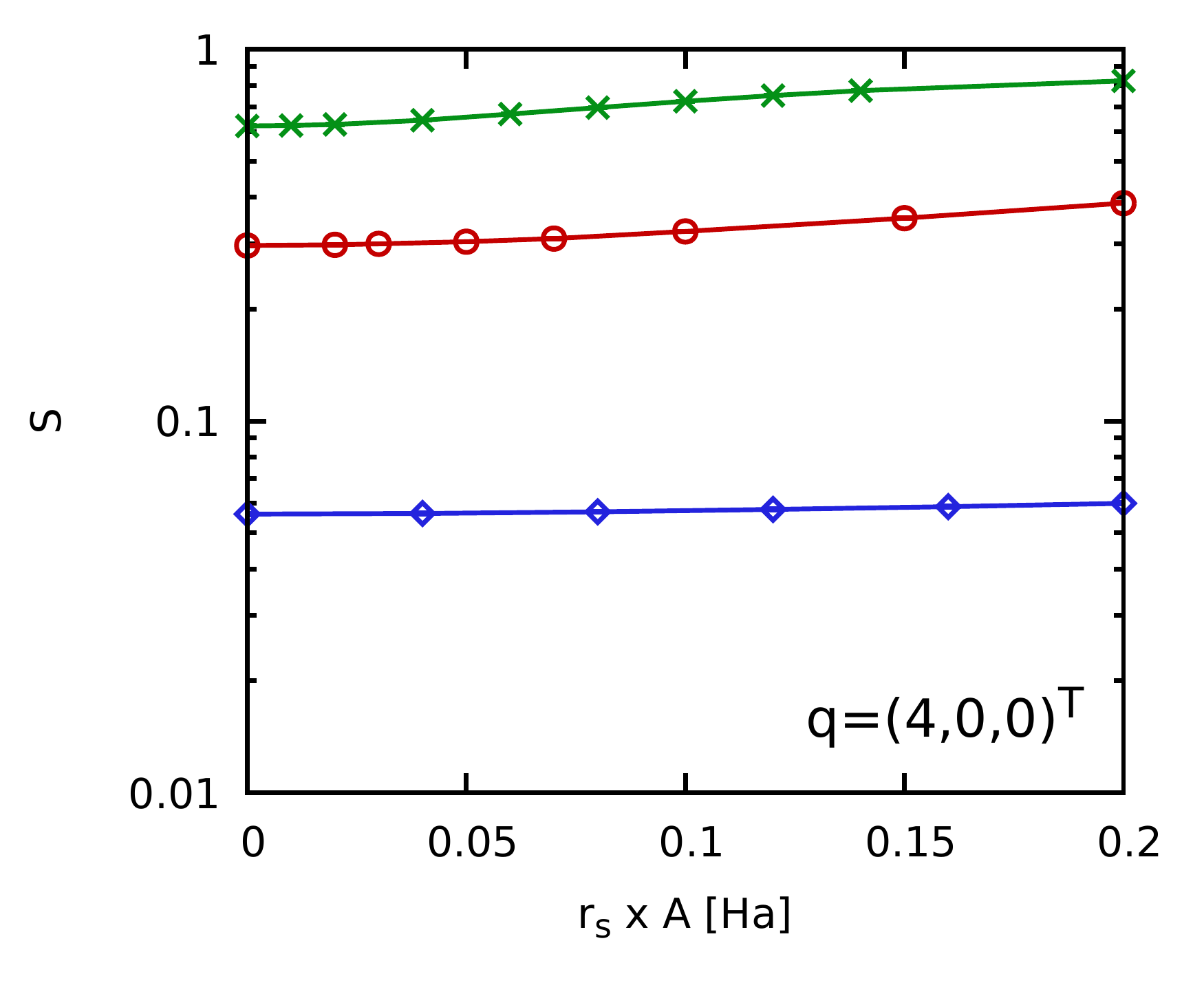}
\caption{\label{fig:N34_sign}
Average sign $S$ (see also Ref.~\cite{dornheim_sign_problem}) in our PIMC simulations of the UEG at $\Theta=1$ and $r_s=4$ (blue diamonds), $r_s=10$ (red circles), and $r_s=20$ (green crosses) as a function of the perturbation strength $A$ for $q=1.27q_\textnormal{F}$ (top) and $q=2.54q_\textnormal{F}$ (bottom).}
\end{figure} 

Let us postpone the discussion of the impact of the coupling parameter on the energy response and briefly touch upon the manifestation of the fermion sign problem on our PIMC simulations. In Fig.~\ref{fig:N34_sign}, we show PIMC data for the average sign $S$, which constitutes a straightforward measure for the amount of cancellation of positive and negative contributions to the fermionic partition function. It is, in fact, well-known~\cite{dornheim_sign_problem} that Monte Carlo error of any observable $O$ is inversely proportional to $S$,
\begin{eqnarray}\label{eq:relative_error}
\frac{\Delta O}{O} \sim \frac{1}{S\sqrt{N_\textnormal{MC}}}\ ,
\end{eqnarray}
where $N_\textnormal{MC}$ is the number of Monte Carlo samples.
This is the origin of the notorious fermion sign problem, which leads to a drastic increase in compute time for small $S$.

First and foremost, we find that the average sign of the unperturbed system strongly depends on the density with $S=0.633$ for $r_s=20$, $S=0.298$ for $r_s=10$, and
$S=0.056$ for $r_s=4$. Thus, the PIMC simulations are particularly expensive in the latter case, where we require more than 100 times the number of Monte Carlo samples (and, hence, 100 times the amount of compute time) to achieve the same accuracy as for $r_s=20$, see Eq.~(\ref{eq:relative_error}). From a physical perspective, this is a direct consequence of the increased coupling strength at low densities, which effectively separates the paths of individual electrons in the PIMC simulation. Therefore, the formation of permutation cycles~\cite{Dornheim_permutation_cycles}, which are exclusively responsible for a non-unity sign in standard PIMC, is suppressed, which explains the observed trend. 

While the average sign by itself does not constitute a physical observable, it can nevertheless give insights into the physical behaviour of a given system in some situations~\cite{Mondaini_Science_2022,https://doi.org/10.48550/arxiv.2207.09026}. In the top panel of Fig.~\ref{fig:N34_sign}, we show the dependence of the average sign on the perturbation amplitude $A$ for $q=1.27q_\textnormal{F}$. Evidently, the sign systematically drops with $A$ for all three depicted densities.
In stark contrast, the sign increases with $A$ for $q=2.54q_\textnormal{F}$ (bottom panel), although the effect is less pronounced for $r_s=4$. This observation can directly be traced back to the basic geometry and length scale of the applied perturbation. For $\mathbf{q}=(2,0,0)^T2\pi/L$, the wave length of the external potential is significantly larger than the average interparticle distance, $\lambda>d$. Therefore, multiple electrons have to occupy the same minimum of $\phi_\textnormal{ext}$. This, in turn, leads to an increase in the degree of quantum degeneracy and, therefore, a decrease in the average sign $S$. In contrast, the electrons are effectively separated by the external potential for $\mathbf{q}=(4,0,0)^T2\pi/L$, which leads to a (small) suppression of the formation of permutation cycles in this regime.

\begin{figure}\centering
\hspace*{0.012\textwidth}\includegraphics[width=0.485\textwidth]{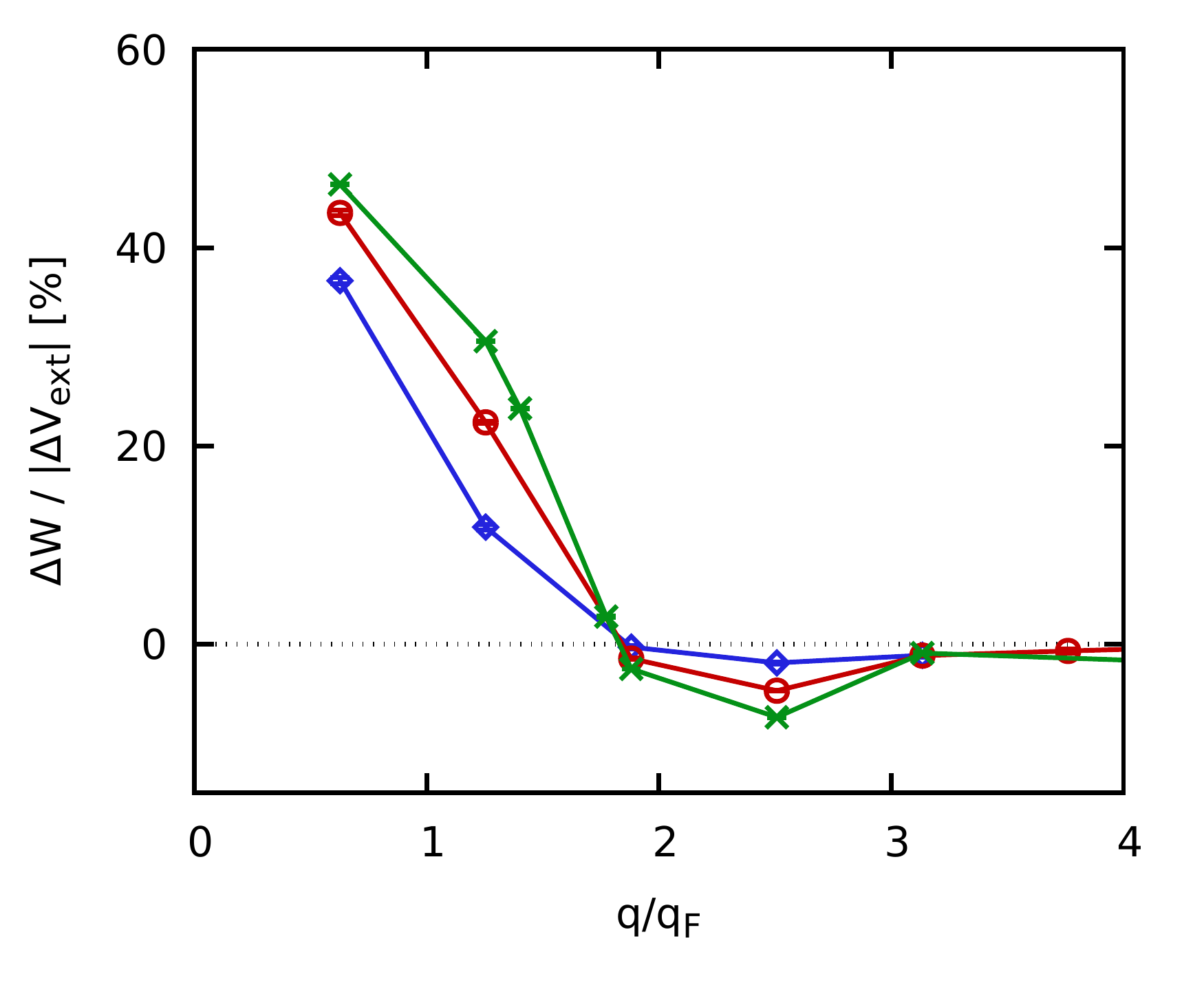}\\\vspace*{-1.28cm}\includegraphics[width=0.5\textwidth]{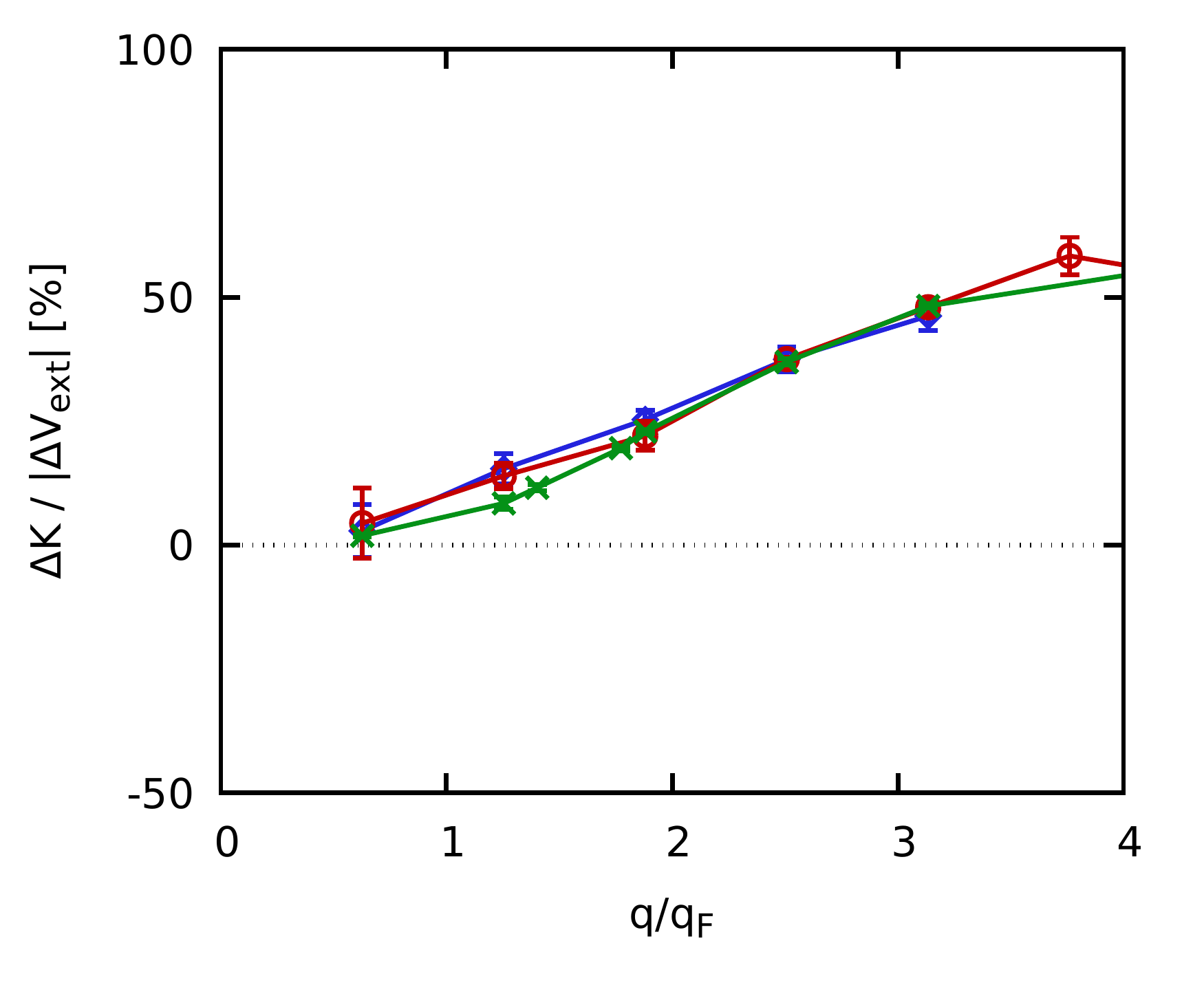}
\caption{\label{fig:N34_compare_rs}
Induced interaction energy change (top) and kinetic energy change (bottom) relative to the external potential energy $V_\textnormal{ext}$ in the linear response limit at $\Theta=1$ for $r_s=4$ (blue diamonds), $r_s=10$ (red circles), and $r_s=20$ (green crosses).}
\end{figure} 

Let us next return to the topic at hand, which is the analysis of the energy response for different values of the density parameter $r_s$. This is shown in Fig.~\ref{fig:N34_compare_rs}, where we divide the linear-response coefficients of the interaction energy $W$ (top) and kinetic energy $K$ (bottom) by the respective coefficient of the external potential energy $V_\textnormal{ext}$ to achieve direct comparability between the different $r_s$. The wave-number dependence of the interaction energy response is mostly dominated by two main trends: 1) for small wave numbers, the interaction energy approaches an increasing fraction of the external potential energy, which can be viewed as a consequence of perfect screening in the UEG; 2) we can resolve the negative minimum around $q=2.5q_\textnormal{F}$ due to the spatial alignment for all three densities, and the effect becomes substantially more pronounced with increasing coupling strength, as it is expected. Finally, the interaction energy becomes rapidly unaffected by the external potential in the single-particle limit, which begins around $q\gtrsim3q_\textnormal{F}$. 
Interestingly, the ratio of the kinetic to the external potential energy monotonically increases with $q$, but seems to be independent of $r_s$ within the given uncertainty interval.

\section{Summary and Discussion\label{sec:summary}}

In this work, we have presented extensive new \emph{ab initio} PIMC results for the energy response of the finite temperature UEG to an external harmonic perturbation. We have found excellent agreement between our data and the scaling predicted by the generalised stiffness theorem both in the regimes of linear and nonlinear response theory. Our investigation has revealed a number of interesting trends in the induced change of different contributions to the total energy budget. A particularly interesting finding is given by the nontrivial change of the interaction energy $\Delta W$, which changes its sign becoming negative for intermediate perturbation wave numbers $q$. This trend can be readily explained in terms of the relevant length scales, and it further substantiates the pair alignment model proposed in Ref.~\cite{Dornheim_Nature_2022}. Furthermore, we have found that this exchange--correlation induced trend becomes less pronounced upon increasing the temperature or decreasing the coupling strength, as it is expected.

Our new results further complete our understanding of the UEG~\cite{status,review}, which is important in its own right. All PIMC results are freely available online~\cite{online_note} and can be used as a rigorous benchmark for the development of new methods and approximations.

Future extensions of our work may include the in-depth analysis of nonlinear effects, and the application of the present framework to realistic systems beyond the UEG model, including WDM~\cite{Bohme_PRL_2022,Bohme_PRE_2023} and other quantum many-body systems such as ultracold atoms~\cite{Dornheim_SciRep_2022,Filinov_PRA_2012}.

\appendix

\section{External potential energy expansion in terms of linear and nonlinear density responses}\label{sec:appendixA}

The external potential energy for the harmonically perturbed electron system, see Eq.~(\ref{eq:Hamiltonian_modified}), is given by
\begin{equation*}
V_{\mathrm{UEG}}^{\boldsymbol{q},A}=\int_{V}\delta{n}(\boldsymbol{r})V_{\mathrm{ext}}(\boldsymbol{r})d^3r\,,
\end{equation*}
with $V_{\mathrm{ext}}(\boldsymbol{r})$ the external potential energy perturbation, $\delta{n}(\boldsymbol{r})$ the perturbed density. Employing the Plancherel theorem and substituting for the Fourier transform of the potential energy perturbation, we have
\begin{align}
V_{\mathrm{UEG}}^{\boldsymbol{q},A}&=\frac{1}{V}\sum_{\boldsymbol{k}}\langle\hat\rho_{\boldsymbol{k}}\rangle_{\boldsymbol{q},A}V_{\mathrm{ext}}(\boldsymbol{k})\Rightarrow\nonumber\\
V_{\mathrm{UEG}}^{\boldsymbol{q},A}&=\frac{N}{V}A\sum_{\boldsymbol{k}}\langle\hat\rho_{\boldsymbol{k}}\rangle_{\boldsymbol{q},A}\left(\delta_{\boldsymbol{k},\boldsymbol{q}}+\delta_{\boldsymbol{k},-\boldsymbol{q}}\right)\Rightarrow\nonumber\\
V_{\mathrm{UEG}}^{\boldsymbol{q},A}&=2\frac{N}{V}A\langle\hat\rho_{\boldsymbol{q}}\rangle_{\boldsymbol{q},A}\,.\nonumber
\end{align}
The non-linear expression for the perturbed density at the first harmonic, $\langle\hat\rho_{\boldsymbol{q}}\rangle_{\boldsymbol{q},A}=\sum_{m=1}^{\infty}A^{2m-1}\chi^{(1,2m-1)}(\boldsymbol{q})$ with $\chi^{(l,m)}(\boldsymbol{q})$ the $m-$th non-linear static response at the $l-$th harmonic, only contains odd powers of the perturbation strength. Substitution leads to the series expansion in orders of the perturbation strength
\begin{equation*}
V_{\mathrm{UEG}}^{\boldsymbol{q},A}=2n\sum_{m=1}^{\infty}A^{2m}\chi^{(1,2m-1)}(\boldsymbol{q})\,,
\end{equation*}
which now only contains the even powers of $A$. It is interesting that, even in the non-linear regime, the contributions of the high harmonics $l>1$ to the external potential energy cancel out and, thus, only the non-linear contributions that concern the perturbation harmonic $l=1$ or $\boldsymbol{k}=\boldsymbol{q}$ survive. The first three terms read as
\begin{equation*}
V_{\mathrm{UEG}}^{\boldsymbol{q},A}=2n\left[A^2\chi(\boldsymbol{q})+A^4\chi^{(1,3)}(\boldsymbol{q})+A^6\chi^{(1,5)}(\boldsymbol{q})+\cdots\right].
\end{equation*}
It is apparent that, exactly as specified in Eq.~(\ref{eq:stiffnessexternal}), the linear term reads as
\begin{equation*}
V_{\mathrm{UEG}}^{\boldsymbol{q},A}=2nA^2\chi(\boldsymbol{q})\,.
\end{equation*}

\section{Nonlinear finite temperature density stiffness theorem}\label{sec:appendixB}

Nonlinear extensions to the ground state density stiffness theorem have been mentioned in earlier QMC investigations, where the second order term was also provided without proof~\cite{moroni,moroni2}. In Appendix~\ref{sec:appendixA}, the external potential energy was expanded in terms of the nonlinear density responses. This is a very important result that allows us to revisit the density functional perspective and to minimize the intrinsic free energy with respect to higher-order density perturbations in a sequential manner. As a consequence, the free energy difference is also gradually expanded in terms of the nonlinear density responses. The procedure resembles the determination of nonlocal corrections to the free energy of non-interacting fermions within the framework of extended Thomas-Fermi theory, as reported in Ref.~\cite{Geldart1985}, but the demanding series inversion, that is based on the Lagrange-Burmann formula, is circumvented. Ultimately, one acquires the \emph{nonlinear finite temperature density stiffness theorem} that reads as
\begin{equation*}
\Delta{F}=-n\sum_{m=1}^{\infty}\frac{2m-1}{m}A^{2m}\chi^{(1,2m-1)}(\boldsymbol{q})\,.
\end{equation*}
It is very interesting that, even in the non-linear regime, the contributions from the high harmonics $\boldsymbol{k}=l\boldsymbol{q}$ with $l>1$ to both the free energy change and the external potential energy cancel out and, thus, only the non-linear contributions that concern the exact perturbation harmonic $\boldsymbol{k}=\boldsymbol{q}$ survive. The first three terms read as
\begin{equation*}
\Delta{F}=-n\left[A^{2}\chi^{(1,1)}(\boldsymbol{q})+\frac{3}{2}A^{4}\chi^{(1,3)}(\boldsymbol{q})+\frac{5}{3}A^{6}\chi^{(1,5)}(\boldsymbol{q})\right]\,.
\end{equation*}
It is apparent that, exactly as specified in Eq.~(\ref{eq:stiffness}), the linear term reads as
\begin{equation*}
\Delta{F}=-nA^2\chi(\boldsymbol{q})\,.
\end{equation*}

\section*{Acknowledgments}

This work was partially supported by the Center for Advanced Systems Understanding (CASUS) which is financed by Germany’s Federal Ministry of Education and Research (BMBF) and by the Saxon state government out of the State budget approved by the Saxon State Parliament.
The PIMC calculations were carried out at the Norddeutscher Verbund f\"ur Hoch- und H\"ochstleistungsrechnen (HLRN) under grant shp00026, and on a Bull Cluster at the Center for Information Services and High Performance Computing (ZIH) at Technische Universit\"at Dresden.

\bibliography{bibliography}

\begin{thebibliography}{150}
\bibitem [{onl()}]{online_note}%
  \BibitemOpen
  \href@noop {} {}\bibinfo {note} {A repository containing all PIMC simulation
  data will be made available upon publication.}\BibitemShut {Stop}%
\end{thebibliography}
\end{document}